\documentclass[aps,physrev, twocolumn, superscriptaddress,amsmath,amssymb,longbibliography,nobibnotes,secnumarabic]{revtex4-2}

\usepackage{graphicx}
\usepackage{dcolumn}
\usepackage{bm}
\usepackage{amsthm}
\usepackage{xcolor}
\usepackage[normalem]{ulem}
\usepackage{xr-hyper}

\usepackage{titletoc}
\DeclareMathOperator{\sech}{sech}

\IfFileExists{Paper1_PRR/References.bib}{
    \newcommand{\paperroot}{Paper1_PRR/}
}{
    \newcommand{\paperroot}{}
}
\newcommand{\taupred}{\tau_\mathfrak{p}}
\newcommand{\tpred}{\tau_\mathfrak{p}}

\newcommand{\tausigma}{\tau_\sigma}
\titlecontents{lsection}
  [0em]
  {\addvspace{0.75em}}
  {\contentslabel{3.0em}}
  {}
  {\hfill\contentspage}

\titlecontents{lsubsection}
  [3.0em]
  {\addvspace{0.12em}}
  {\contentslabel{3.7em}}
  {}
  {\hfill\contentspage}

\titlecontents{lsubsubsection}
  [6.7em]
  {}
  {\contentslabel{1.8em}}
  {}
  {\hfill\contentspage}
\usepackage[unicode=true, psdextra, hidelinks]{hyperref}
\begin{document}

\title{Predictable Mean-Field Chaos in Random Recurrent Neural Networks}

\author{Alkesh Yadav}
\thanks{These authors contributed equally to this work.}
\affiliation{Edmond and Lily Safra Center for Brain Sciences, The Hebrew University of Jerusalem, 9190401, Israel}

\author{Vladimir Shaidurov}
\thanks{These authors contributed equally to this work.}
\affiliation{Edmond and Lily Safra Center for Brain Sciences, The Hebrew University of Jerusalem, 9190401, Israel}

\author{Jonathan Kadmon}
\email[]{jonathan.kadmon@mail.huji.ac.il}
\affiliation{Edmond and Lily Safra Center for Brain Sciences, The Hebrew University of Jerusalem, 9190401, Israel}

\date{\today} 

\begin{abstract}
Dynamical mean-field theory (DMFT) maps deterministic chaos in random recurrent neural networks to an effective Gaussian process, usually treated as an ensemble description rather than a predictor of individual trajectories. Contrary to this view, we show that the chaotic mean-field process of the Sompolinsky--Crisanti--Sommers model can be perfectly predictable at the level of a single neuron. When the nonlinearity has Gaussian or faster Fourier decay, the exact continuous past of one realization determines its entire future: the conditional prediction error vanishes despite a positive Lyapunov exponent. We trace this result to complex-time singularities of the self-consistent covariance through the Paley--Wiener criterion. A Lanczos/Krylov representation turns the covariance into a state-space predictor and defines $\alpha$, the rate at which predictive information is transferred to higher temporal modes. For smooth odd nonlinearities in this predictable class, the Krylov rate and the largest Lyapunov exponent scale differently near the chaotic transition, demonstrating that predictive complexity and microscopic instability are distinct characteristics of the dynamics. Resolving the first $p$ Krylov modes yields a prediction horizon that grows as $\log p$ and enables finite-network forecasts from a single neuron's observed past, without knowing the connectivity and without observing the rest of the network. Thus, the mean-field power spectrum becomes a diagnostic of whether apparent variability reflects irreducible fluctuations or hidden deterministic structure encoded in the observed past.
\end{abstract}

\maketitle

\section{Introduction}
Large recurrent networks of randomly coupled rate neurons are deterministic many-body systems, yet beyond a critical coupling they generate high-dimensional chaos~\cite{SCS1988}. In the thermodynamic limit, dynamical mean-field theory (DMFT) replaces the network by a single-site \emph{stochastic} process: a typical neuron follows a stationary Gaussian law with a self-consistent covariance kernel~\cite{SCS1988,kadmon2015}. The reduced description is ordinarily understood as an ensemble theory rather than a predictor of individual trajectories. Once the microscopic connectivity and the instantaneous states of the remaining neurons have been eliminated, the history of one neuron appears insufficient to determine its future. The effective Gaussian drive is therefore treated as genuine temporal noise, continually introducing uncertainty that cannot be removed by observing that neuron alone.

Whether this effective noise is an irreducible source of new information is not decided by the usual characterization of chaos through Lyapunov exponents and decorrelation times. Classical prediction theory makes the relevant distinction precise. Wiener, Kolmogorov, and Krein classified stationary Gaussian processes as \emph{regular}, continually generating innovations that no past can anticipate, or \emph{singular}, with the continuous past uniquely determining the future; the distinction is fixed by a log-integrability condition on the power spectrum~\cite{yaglom1962introduction,DymMcKean1976}. A positive largest Lyapunov exponent, $\lambda_L$, might seem to place the chaotic mean-field process on the regular side. However, $\lambda_L$ measures the divergence of nearby microscopic network states under perturbation, whereas singularity measures the conditional uncertainty of one realized mean-field trajectory given its exact history. These notions are logically independent. Nor is predictability determined by the decorrelation time: an Ornstein--Uhlenbeck process can retain correlations for arbitrarily long times while remaining innovation-driven, whereas an analytic process can decorrelate on an $\mathcal{O}(1)$ timescale while encoding its entire future in its past.

In this Letter we show that the chaotic mean-field process of the paradigmatic Sompolinsky--Crisanti--Sommers (SCS) network~\cite{SCS1988} can, contrary to the standard expectation, be singular. Which prediction-theoretic class it belongs to is decided by the analytic structure of the single-neuron nonlinearity. For nonlinearities with Gaussian or faster Fourier decay, such as $\operatorname{erf}$, the effective process remains random across realizations but generates no new uncertainty after conditioning on the exact continuous past; for the canonical $\tanh$, infinitely smooth but with slower Fourier decay, innovations persist. In the singular class, the self-averaging covariance kernel supplies a universal continuation rule, while the observed history selects the particular trajectory to be continued. Thus, without knowing the microscopic connectivity or observing the rest of the network, the future of a typical neuron is fixed by its own continuous past. Information that appears to have been lost in the mean-field reduction survives, encoded in the temporal structure of the trajectory.

Perfect predictability, however, does not explain how this information is organized or how much of it must be resolved in practice. Analyticity formally encodes the future in an infinite hierarchy of time derivatives, but raw derivatives form a poorly conditioned basis and provide no natural measure of the cost of prediction at finite resolution. The Lanczos/Krylov construction~\cite{Viswanath_Recursion,Parker,Rabinovici2025KrylovComplexity} supplies such a representation by unfolding the mean-field power spectrum into a semi-infinite chain of orthogonal temporal modes~\cite{teschl2000jacobi}. The resulting Jacobi operator turns the covariance kernel into a state-space predictor, while the asymptotic growth of its Lanczos coefficients quantifies how rapidly the information required for prediction is transferred to higher-order modes. This defines a predictive complexity rate that can be compared with the Lyapunov instability of the underlying network.

For the SCS model we find that this complexity rate, $\alpha$, exceeds $\lambda_L$ across the nonlinearities and parameter ranges studied. At finite resolution, the construction becomes a practical forecasting scheme, and resolving the first $p$ Krylov modes extends the prediction horizon as $\log p$. Microscopic instability and predictive complexity thus characterize distinct aspects of mean-field chaos.

\section{Predictability of SCS Mean-Field Chaos}
\label{sec:mft}

We consider the Sompolinsky--Crisanti--Sommers (SCS) model of $N$ continuous variables with random couplings, $J_{ij}\sim\mathcal{N}(0,g^2/N)$,
\begin{equation}
\dot x_i(t) = -x_i(t) + \sum_{j=1}^{N} J_{ij}\,\phi(x_j(t)).
\label{eq:scs}
\end{equation}
In the thermodynamic limit, DMFT reduces the network to an ensemble of independent single-unit dynamics, $\dot x = -x + \eta(t)$, with $\eta(t)$ a stationary Gaussian process~\cite{SCS1988}. For $g>g_c$, the fixed point destabilizes and the system enters a chaotic phase. The stationary autocorrelation $\Delta(\tau)=\langle x(t)x(t+\tau)\rangle$ obeys a Newtonian equation of motion, $\ddot{\Delta}(\tau) = -\partial_\Delta V(\Delta)$, in an effective potential,
\begin{equation}
\begin{aligned}
    V(\Delta) & = -\tfrac{1}{2}\Delta^{2} + g^{2}\,\langle \Phi(x)\Phi(x')\rangle_{\Delta},
\label{eq:newton}
\end{aligned}
\end{equation}
with
$\Phi'(x)=\phi(x)$, initial conditions $\Delta(0)=\Delta_{0}$, $\dot\Delta(0)=0$, and $\langle\cdot\rangle_\Delta$ denoting an average over $x,x'$ jointly Gaussian with variance $\Delta_{0}$ and covariance $\Delta$ \cite{kadmon2015}.

We now formalize the conditional trajectory question posed above. Once spatial degrees of freedom have self-averaged, single-trajectory prediction is governed by the stationary covariance $\Delta(\tau)$ and by its Fourier transform, the power spectral density $\tilde{\Delta}(\omega)=\int d\tau\,e^{i\omega\tau}\Delta(\tau)$. These objects determine whether the observed past of one realization contains all the information needed to reconstruct its future.

\paragraph*{Predictability of GP---}
For a Gaussian process with a continuous past $X(u\le 0)$, the optimal estimator $\hat X(t)=\mathbb{E}[X(t)\mid X(u\le 0)]$ minimizes the mean-square error. We measure global accuracy by the Laplace-weighted cumulative error
\begin{equation}
    \sigma_{\infty}^{2}(\taupred) := \frac{2}{\taupred}\int_{0}^{\infty}\mathbb{E}\!\big[ (X(t)-\hat X(t))^2  \big] e^{- 2t/\taupred}\,dt.
\label{eq:sigmainf}
\end{equation}
Because the integrand is non-negative, $\sigma_{\infty}^{2}(\taupred)=0$ characterizes \emph{perfect predictability}---the future is uniquely determined by the past.
This time-domain prediction error has an exact spectral representation in terms of $\tilde{\Delta}(\omega)$ (Sec. S1 of~\cite{SM}):
\begin{equation*}
     \sigma^2_\infty(\taupred) = \frac{1}{2\pi}\int_{-\infty}^{\infty} d\omega' \exp \left( {\frac{1}{\pi\taupred}\int_{-\infty}^{\infty}\frac{\ln \tilde{\Delta}(\omega + \omega')}{\taupred^{-2}+\omega^2} \, d\omega }\right).
     \label{eq:yaglom_error}
\end{equation*}
Based on the analytic properties of causal systems~\cite{kubo1991statistical, yaglom1962introduction}, the power spectral density can be factorized as $\tilde{\Delta}(\omega) = |\chi(i \omega)|^2$, where $\chi(i\omega)$ represents a causal linear response function that generates the observed stochastic process from an underlying delta-correlated noise source. The exponential term 
corresponds to the analytic continuation of this response function into the complex frequency plane, yielding $|\chi(i\omega + 1/\taupred)|^2$ for $\taupred>0$. Physically, this evaluates the variance of the system's irreducible fluctuations---the fundamental, uncorrelated stochastic degrees of freedom driving the dynamics---at a given frequency scale. The outer integral then
aggregates these uncorrelated innovations to yield the total forecast error.

Consequently, the process is fully deterministic if and only if the argument of the exponential diverges to negative infinity, which precludes the existence of the response function, $\chi(i \omega)$. This exactly yields the celebrated Paley--Wiener (or Krein) criterion for process singularity~\cite{DymMcKean1976}:
\begin{equation} 
    \int_{-\infty}^{\infty} \frac{\ln \tilde{\Delta}(\omega)}{\taupred^{-2} + \omega^2} \, d\omega = -\infty \iff \sigma^2_{\infty} = 0.
    \label{eq:krein}
\end{equation}
A PSD with an exponential or faster tail makes the integrand in~\eqref{eq:krein}
diverge, resulting in a completely predictable process.
In the following section, we demonstrate that the PSD tail of the chaotic GP is critically sensitive to the analytic properties of the activation function, $\phi(x)$. We derive this by performing an analytic continuation of the auto-correlation $\Delta(\tau)$ into the complex $\tau$-plane. 

\paragraph*{PSD tail of $\Delta(\tau)$.---}

The link between the microscopic network and its macroscopic predictability is established through a strict analytic chain: the high-frequency Fourier decay of the activation function $\phi(x)$ dictates the radius of convergence of the effective potential $V(\Delta)$, which in turn fixes the exact complex-time singularity of the autocorrelation $\Delta(\tau)$ and the PSD tail.

By expanding the primitive $\Phi(x)$ in a basis of orthonormal Hermite polynomials, we obtain a Taylor series for the potential: $V(\Delta) = -\Delta^2/2 + g^2\sum_{n=0}^\infty (A_n^2/n!) \Delta^n$. The expansion coefficients are given by $A_n = i^n/(2\pi) \int_{-\infty}^\infty \hat{\Phi}(k) k^n e^{-\Delta_0 k^2/2} \, dk$, where $\hat{\Phi}(k)$ is the Fourier transform of $\Phi(x)$. The analytic structure of $V(\Delta)$, dictated by its radius of convergence $S$, is strictly governed by the asymptotic decay rate of $\hat{\Phi}(k)$.

For nonlinearities with Gaussian (or faster) Fourier decay, e.g. $\phi(x) = \operatorname{erf}(x/\sqrt{2}\sigma)$
with 
$\hat\Phi(k)\sim k^{-2}\,\exp({-(k^{2}/2) \sigma^2})$, 
the radius of convergence is given by the singularity at $|\Delta|=S$, an \emph{algebraic branch point} arising from the divergence of the Gaussian expectation $\langle \Phi\Phi'\rangle_{\Delta}$.
The rapid frequency-domain decay over-compensates the divergence of the Gaussian measure, pushing the branch point to $S= \Delta_{0} + \sigma^2$, strictly \emph{outside} the physical region of $|\Delta| \leq \Delta_0$ (Sec. S4 of~\cite{SM}). The potential is therefore analytic throughout the physically accessible domain. For nonlinearities with faster than Gaussian Fourier decay (e.g., polynomials), $V(\Delta)$ is entire and the singularities are pushed to infinity.

\begin{figure}[t]
\begin{center}
    \includegraphics[width=1\columnwidth]{\paperroot 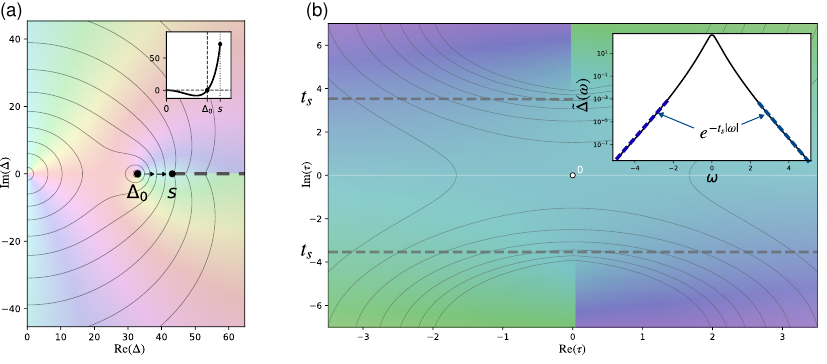}
\end{center}
\caption{\textbf{Analytic structure of the smooth chaotic mean-field process.} \textbf{(a)} Analytic continuation of the effective potential $V(\Delta)$ for $\phi(x)=\operatorname{erf}(x)$. Color denotes phase and contours denote magnitude; the nearest branch point lies at $\Delta=S>\Delta_0$, outside the physical interval of autocorrelations. Inset: $V(\Delta)$ on the real axis. \textbf{(b)} Analytic continuation of the autocorrelation $\Delta(\tau)$ in complex time. The dashed lines mark the analytic strip $|\operatorname{Im}\tau|<t_S$. Inset: the PSD $\tilde{\Delta}(\omega)$ on a semilog scale, showing the exponential tail set by the singularity at $t_S$.}
\label{fig:analytic}

\end{figure}

Energy conservation gives the time-of-flight integral
\begin{equation}
\tau(\Delta) = \int_{\Delta_{0}}^{\Delta}\frac{dz}{\sqrt{2\,[V(\Delta_{0})-V(z)]}}.
\label{eq:tof}
\end{equation}
Since $\Delta_{0}$ is a regular point, $\Delta(\tau)$ is locally analytic with $\Delta(\tau)-\Delta_{0}\propto\tau^{2}$. To reach the branch point at $S>\Delta_{0}$ the trajectory must continue past the physical turning point, where the integrand becomes purely imaginary; 
the singularity of $\Delta(\tau)$ therefore sits on the imaginary axis at 
\begin{equation}
t_{S} = \left|\,\int_{\Delta_{0}}^{S}\frac{dz}{\sqrt{2\,[V(\Delta_{0})-V(z)]}}\,\right|,
\label{eq:ts}
\end{equation}
which remains finite even as $S\to\infty$ (Sec. S5 of~\cite{SM}). 
By standard Fourier asymptotic analysis~\cite{paley-weiner}, the high-frequency tail of the PSD is  controlled by the singularity of $\Delta(\tau)$ closest to the real axis in the complex time plane. For activation functions such as $\operatorname{erf}(x)$, this singularity occurs at $\pm i t_S$ (Sec. S5 of~\cite{SM}), yielding the exact asymptotic tail:  
\begin{equation}
    \tilde\Delta(\omega) \,\sim\, \exp(-t_{S}\,\omega), \qquad \omega\to\infty.
\label{eq:psdtail}
\end{equation}
Therefore, the power spectrum decays exponentially, with the decay rate set by a singularity of the potential lying \emph{outside} the physically realizable autocorrelation (Fig.~\ref{fig:analytic}), and the resultant GP is completely predictable.

What matters in the preceding argument is not the particular choice of $\operatorname{erf}$, but the Gaussian or faster Fourier decay of the nonlinearity. Common activation functions such as $\tanh$ and ReLU lie outside this regime. For $\phi=\tanh$, the complex poles give only exponential Fourier decay and pin the singularity of $V(\Delta)$ to the physical boundary $S=\Delta_0$. In this case $V$ and all derivatives remain finite as $\Delta\to\Delta_0^-$, but the Taylor series has zero radius of convergence.
The PSD tail is then stretched exponential, rather than exponential, and the Paley--Wiener criterion fails (Sec. S5 of~\cite{SM}). For ReLU or hard tanh, finite differentiability of $V$ at $\Delta_0$ gives an algebraic PSD tail, so predictability fails outright. Convolving either activation with a Gaussian of width $\sigma$ restores a Gaussian Fourier tail. For every $\sigma>0$ the nearest singularity is displaced outside the physical domain and the process regains the exponential tail~\eqref{eq:psdtail}. As $\sigma\to0$, the original activation is recovered but the analytic strip, and hence the finite-resolution prediction scale, collapses.

We conclude that \emph{for every smooth $\phi$ with Gaussian or faster Fourier tail
---the chaotic mean-field GP belongs to the Paley--Wiener class of singular Gaussian processes}: its future is completely determined by its continuous past, even deep in the chaotic phase. 
This path analyticity is distinct from macroscopic decorrelation. An Ornstein--Uhlenbeck process with long correlation time looks smooth but is nowhere differentiable, whereas erf-driven DMFT can decorrelate on $\mathcal{O}(1)$ timescales yet generate strictly analytic paths. 
We now unfold the covariance into a Krylov state space to expose how this latent determinism is organized.

\section{Krylov State Space and Prediction Complexity}
The Paley--Wiener criterion shows that the analytic mean-field process is 
perfectly predictable. 
What remains is to expose the dynamical structure that stores this information. We do so by unfolding the covariance kernel into a Lanczos/Krylov state space, where temporal information is organized into an orthogonal hierarchy of modes evolving under a deterministic linear generator.

Starting from the PSD, we define the spectral measure
\( d\mu(\omega)=\tilde\Delta(\omega)\,d\omega .\) 
Let \(\{\phi_n(\omega)\}_{n\ge0}\) be the corresponding orthonormal polynomials.
Multiplication by frequency is tridiagonal in this basis,
\(
\omega\phi_n(\omega)
=
\alpha_{n+1}\phi_{n+1}(\omega)
+
\alpha_n\phi_{n-1}(\omega),
\)
where $\alpha_n>0$ and the diagonal coefficients vanish by the evenness of the measure. The coefficients \(\alpha_n\) define the semi-infinite Jacobi operator
\begin{equation}
J=\sum_{n\ge0}\alpha_{n+1}
\left(
|e_n\rangle\langle e_{n+1}|
+
|e_{n+1}\rangle\langle e_n|
\right),
\end{equation}
so that multiplication by \(\omega\) is represented as nearest-neighbor hopping on a one-dimensional Krylov chain.

For a realized trajectory \(X(t)\), we define Krylov coordinates
\(
s_n(t)=\phi_n(i\partial_t)X(t),
\:
|s(t)\rangle=\sum_{n\ge0}s_n(t)|e_n\rangle .
\)
The Lanczos recursion gives the exact linear evolution
\begin{equation}
\frac{d}{dt}|s(t)\rangle=-iJ|s(t)\rangle .
\end{equation}
The original trajectory is recovered from the boundary coordinate,
\begin{equation}
X_{\rm K}(t)=\operatorname{Re}\langle e_0|s(t)\rangle .
\end{equation}
Equivalently, the PSD of the mean-field process is the local density of states, $\rho_0(\omega)$, of \(J\) at the boundary site (Sec. S6 of~\cite{SM}),
\begin{equation}
\tilde{\Delta}(\omega)
= \rho_0(\omega) = 
\mathcal{F}_t\!\left[
\langle e_0|e^{-iJt}|e_0\rangle
\right](\omega).
\end{equation}

The initial Krylov state is fixed by the observed past through
\begin{equation}
s_n(0)=\phi_n(i\partial_t)X_{\rm past}(t)\big|_{t=0}.
\end{equation}
Thus, 
the full infinite vector \(|s(0)\rangle\) contains the information needed to continue the trajectory. Once \(|s(0)\rangle\) is specified, the evolution under \(J\) 
gives exact prediction,
\(
\mathbb{E}\!\left[(X(t)-X_{\rm K}(t))^2\right]=0 
\);
the randomness of the Gaussian process appears only through the initial Krylov state selected by the realized past. 

This construction is the orthonormalized version of the local derivative hierarchy. Raw derivatives \(X^{(n)}(0)\) form a poorly conditioned moment basis, since their covariance is determined by the spectral moments of \(\tilde\Delta(\omega)\). The Lanczos construction diagonalizes this covariance: for a Gaussian process, the initial Krylov coordinates \(s_n(0)\) are independent unit-variance Gaussian variables. Finite prediction is therefore not a Taylor truncation, but a controlled projection onto the first \(p\) orthogonal temporal modes.

The chain defines a natural measure of predictive complexity. For spectra with exponential frequency tails, the Lanczos coefficients grow asymptotically linearly~\cite{Lubinsky1988,Parker}, 
\begin{equation}
    \alpha_n \sim \alpha n. 
\end{equation}
This growth rate is strictly set by the distance to the nearest complex-time singularity of the covariance, $\alpha = \pi/(2 t_S)$. 
Physically, $\alpha$ controls the growth of the conventional Krylov complexity, $\sum_n n|\langle e_n|e^{-iJt}|e_0\rangle|^2$, which describes the spreading of a state initially localized at the boundary mode $|e_0\rangle$ into the chain and quantifies how rapidly predictive information is transferred to higher-order temporal modes~\cite{Parker}. A larger $\alpha$ therefore corresponds to faster propagation along the Krylov chain.

Near the critical transition to chaos, the effective potential is quartic ($\phi^4$) for odd nonlinearities, yielding a GP kernel of the form $\Delta(\tau) = \Delta_0 \operatorname{sech}(\tau/\tau_\infty)$, where $\tau_\infty$ is the macroscopic mixing timescale~\cite{CrisantiSompolinsky2018}. Consequently, the singularity sits exactly at $t_S = \pi \tau_\infty / 2$, locking the Krylov growth rate to the inverse mixing time, $\alpha = 1/\tau_\infty$ (Sec. S8 of~\cite{SM}). Near the transition, $\tau_\infty$ diverges as $(g-g_c)^{-1}$~\cite{CrisantiSompolinsky2018}, so $\alpha$ vanishes linearly in $g-g_c$; the largest Lyapunov exponent instead vanishes quadratically, $\lambda_L\sim(g-g_c)^{2}$~\cite{kadmon2015}, making the two rates parametrically distinct at criticality.
Moving into the chaotic phase, however, higher-order corrections to the potential decouple $t_S$ from $\tau_\infty$, breaking this simple proportionality. 
For $g\gg g_c$,
$\alpha$ grows approximately linearly with $g$, to the leading order $t_S \sim g^{-1}$. 
The behavior of smoothed ReLU-like nonlinearities near criticality is governed by a cubic potential ($\phi^3$) but results in a similar scaling of the complex time singularity with the mixing time, $t_S = \pi \tau_\infty$ (Sec. S8 of~\cite{SM}). For $g>g_c$, $\alpha$ exhibits a faster-than-linear growth when the mean connectivity for the ReLU network is held fixed. 

\begin{figure}
    \includegraphics[scale = 0.45]{\paperroot 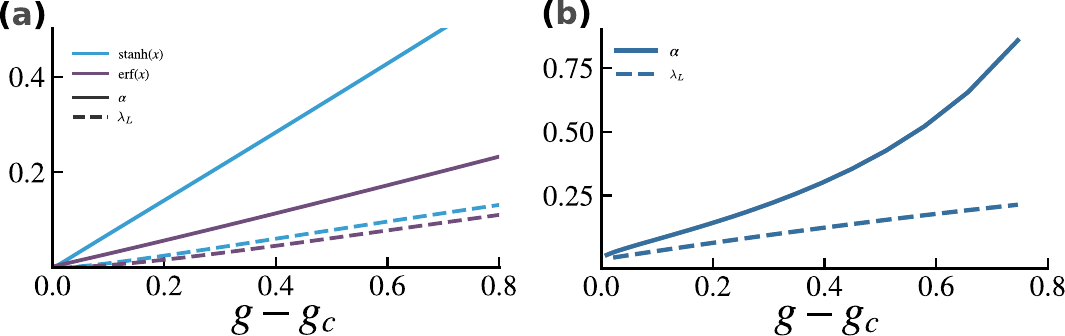}
   \caption{\textbf{Krylov growth}. \textbf{(a)} Krylov growth rate $\alpha$ (solid) and largest Lyapunov exponent $\lambda_L$ (dashed) versus distance from the chaotic transition, $g-g_c$, for smooth odd nonlinearities. Colors denote $\operatorname{erf}(x)$ and smoothed $\tanh(x)$. \textbf{(b)} Same comparison for smoothed ReLU, showing $\alpha>\lambda_L$ over the parameter range investigated.}
   \label{fig:Krylov_growth}
\end{figure}

Figure~\ref{fig:Krylov_growth} compares the Krylov growth rate $\alpha$ with the largest Lyapunov exponent $\lambda_L$ of the underlying network~\cite{EngelkenWolfAbbott2023}. For erf, smoothed tanh, and smoothed ReLU nonlinearities, the DMFT-computed rate $\alpha=\pi/(2t_S)$ is finite and grows with disorder strength, reflecting the flow of predictive information into higher temporal modes. Removing the regularization collapses the analytic strip, leaving $\lambda_L$ finite while $\alpha$ diverges as $\sigma\to0$ (Sec. S7 of~\cite{SM}). Across the regularized nonlinearities and parameter ranges shown, we find $\alpha>\lambda_L$. For smooth odd nonlinearities this ordering follows analytically sufficiently close to the chaotic transition, whereas away from the transition and for smoothed ReLU it is a numerical observation (Sec. S8 of~\cite{SM}). Parker \textit{et al.} derived the relation $\lambda_L\leq2\alpha$ between Lyapunov and Lanczos growth rates in Hamiltonian operator dynamics~\cite{Parker}. Here the Jacobi operator is constructed from the DMFT covariance rather than from a microscopic Liouvillian, so that relation does not directly apply. The stronger ordering $\alpha>\lambda_L$ observed in the networks studied here therefore suggests, but does not establish, an analogous constraint in non-Hamiltonian mean-field dynamics.

\section{Finite resolution prediction}

Exact prediction in the analytic regime requires the full infinite Krylov state. At finite temporal resolution, only the first $p$ Krylov coordinates are resolved, and the remaining modes become a source of conditional uncertainty. We therefore split
\begin{equation}
|s(t)\rangle = |h(t)\rangle \oplus |\xi(t)\rangle, \qquad |h(t)\rangle = \sum_{n=0}^{p-1} s_n(t)|e_n\rangle .
\end{equation}
This head--tail decomposition is the Krylov analogue of a Mori--Zwanzig~\cite{Dynamical_correlations} projection: the resolved modes form the observed subsystem, while the unresolved tail, commonly referred to as a terminator \cite{Haydock_1985,Muck_2022},  acts as a bath whose influence can be integrated out and then approximated by a finite Markovian closure \cite{Viswanath_Recursion}.

The Krylov modes are the orthonormalized form of the local derivative hierarchy, so for the Gaussian mean-field process the initial coordinates $s_n(0)$ are independent unit-variance Gaussian variables. Conditioning on $|h(0)\rangle$ therefore leaves $|\xi(0)\rangle$ as an independent Gaussian field.

Finite-resolution prediction is thus a conditional-inference problem. The full Krylov dynamics are deterministic; uncertainty enters only through the unresolved initial tail. The optimal forecast is the conditional mean over tail realizations, and the prediction error is the corresponding conditional variance.

With this decomposition, the Jacobi operator takes the block form
\begin{equation}
J =
\begin{pmatrix}
J_{hh} & \alpha_p |e_{p-1}\rangle\langle e_p|\\
\alpha_p |e_p\rangle\langle e_{p-1}| & J_{\xi\xi}
\end{pmatrix}.
\end{equation}

The unresolved tail affects the resolved chain only through the boundary coupling $\alpha_p$ and the tail dynamics $J_{\xi\xi}$, allowing it to be integrated out as a self-energy correction to the finite Krylov predictor.

To isolate the dynamics accessible from the resolved head, we eliminate the tail in the Fourier--Laplace domain. This gives an exact finite-dimensional representation for the forecasted boundary coordinate,
\begin{equation}
    \tilde{X}_K(z) = i\langle e_0| [z - J_{hh} - \Sigma_p(z)P_{p-1}]^{-1} |h(0)\rangle.
\end{equation}
Here  $P_{p-1} = |e_{p-1}\rangle\langle e_{p-1}|$ is the boundary projection, and
\begin{equation}
\Sigma_p(z) = \alpha_p^2\int_{\mathbb{R}} \frac{d \rho_p(\omega)}{z - \omega} = \alpha_p^2\sum_{k \geq 0} \frac{\langle \omega^k \rangle}{z^{k+1}} 
\end{equation}
where $\langle \omega^k \rangle = \int \omega^k d\rho_p(\omega) = \langle e_p | J_{\xi\xi}^k|e_p \rangle$ are the moments of the LDOS at $p$, $\rho_p(\omega)$. 
Because the matrix $J$ has zero on the diagonal, 
all odd moments are zero and the expansion of $\Sigma_p(z)$ contains only odd terms in $z$.

We approximate the unresolved high-frequency tail, described by the LDOS $\rho_p(\omega)$, as an Ornstein--Uhlenbeck (OU) process with PSD, $\tilde{\Delta}^{\rm app}(\omega) = D/(\omega^2+ \omega_0^2)$, which corresponds to treating the deep Krylov modes as a Markovian thermal bath with a Lorentzian spectral density.
The corresponding resolvent is given by 
\begin{equation*}
\Sigma^{\rm app}(z) = \frac{\pi D}{\omega_0} \left( \frac{1}{z + i\omega_0} \right) = \frac{\pi D}{\omega_0} \left( \frac{1}{z} - \frac{i\omega_0}{z^2} - \frac{\omega_0^2}{z^3} +  
\dots \right) 
\end{equation*}
We match terms of $\Sigma_p(z)$ to the real part of $\Sigma^{\rm app}(z)$, which only has odd terms in $z$, to obtain the value of parameters $D$ and $\omega_0$.

Thus the finite bath parameters are fixed by the Lanczos coefficients adjacent to the resolved head. Higher-order closures follow by 
constructing higher order approximations of the LDOS, $\rho_p(\omega)$.

\begin{figure}[t]
    \includegraphics[scale = 0.29]{\paperroot 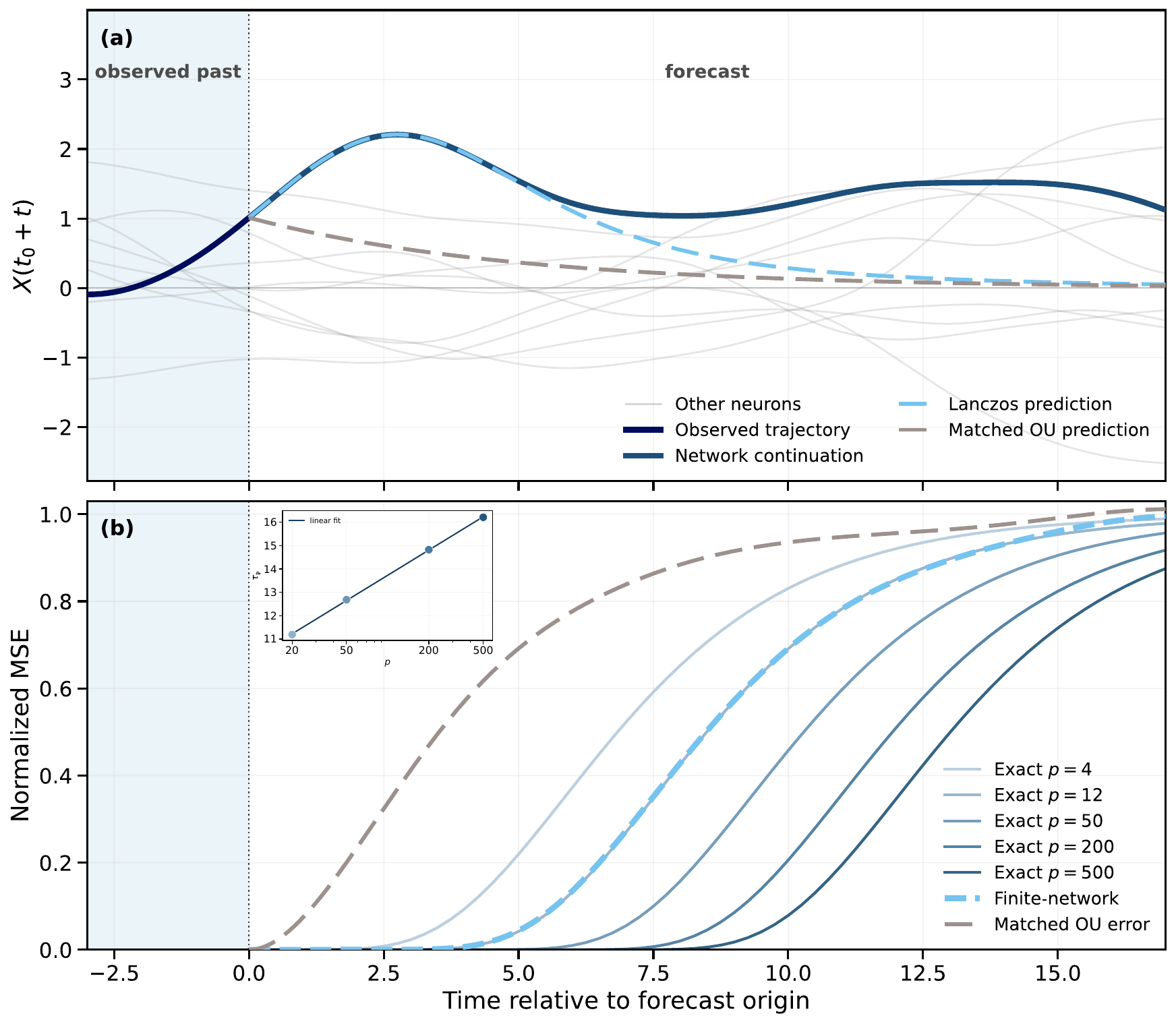}
    \caption{\textbf{Conditional inference of future from the observed past at finite resolution.}
    \textbf{(a)} Forecast of a single-neuron trajectory generated by an SCS network with $N=10^4$ neurons, gain $g=1.5$, and activation $\phi(x)=\operatorname{erf}(x)$.
    The solid and dashed blue curves show the network continuation and Lanczos
    prediction; the gray dashed curve shows the OU prediction with the
    same asymptotic correlation time $\tau_\infty$.
    \textbf{(b)} Normalized conditional prediction error versus time for increasing Krylov resolution $p$. Solid blue curves show the exact mean-field errors. 
    Dashed blue and gray curves show the empirical MSE of the finite-network Lanczos prediction, 
    and the error of the matched OU process.  \textbf{Inset:} the prediction horizon $\taupred^*$, defined 
    by $\sigma_\infty^2(p,\taupred^*)=\varepsilon$,
    grows logarithmically with $p$. 
}
\label{fig:Ou_tail}
\end{figure}

The resulting finite-resolution forecast is the conditional mean over the unresolved initial tail, $\bar X_K(t)=\mathbb{E}_{\xi}[X_K(t)\mid h(0)]$. We quantify its uncertainty by the conditional variance
\begin{equation}
\mathcal{E}_p(t)=
\mathbb{E}_{\xi}
\left[
\left(
X_K(t)-\bar X_K(t)
\right)^2
\mid h(0)
\right].
\end{equation}
Figure~\ref{fig:Ou_tail} illustrates this construction for an SCS network with nonlinearity, $\phi(x) =\operatorname{erf}(x)$, and disorder strength, $g=1.5$.
As $p$ increases, 
the conditional variance remains small for longer times, and the prediction horizon grows. 
Comparison of the Krylov prediction with an OU process matched in decorrelation time, $\tau_\infty$, highlights the distinctive predictability of the chaotic Gaussian process: its observed derivatives, encoded by the finite Krylov modes, strongly constrain its future evolution.
A WKB analysis (Sec. S9 of Ref.~\cite{SM}) shows that the longest time over which the prediction error $\sigma_\infty(\taupred)$ remains small asymptotically scales as
\begin{equation}
\frac{\taupred}{t_S}\sim \log p.
\end{equation}
Near the critical point, where $t_S=\pi\tau_\infty/2$, this becomes $\taupred/\tau_\infty\sim\log p$, demonstrating that exact predictability is recovered logarithmically with increasing observational resolution $p$.

\section{Conclusions}
The stochastic process produced by DMFT need not be merely an ensemble description. In the smooth chaotic SCS model, its covariance kernel also contains the information needed to predict individual mean-field trajectories. After the spatial degrees of freedom self-average, the remaining temporal process is not, in general, an innovation process---one whose present is unpredictable from its own past. For analytic nonlinearities with sufficiently fast Fourier decay, it is a singular Gaussian process whose continuous past fixes its future. This predictability is distinct from Lyapunov instability. While $\lambda_L$ measures sensitivity to microscopic perturbations, the Krylov rate $\alpha$ measures how rapidly the temporal information needed for reconstruction is pushed into higher modes\cite{Rabinovici2025KrylovComplexity,DymarskySmolkin2021,Chapman2025}. We exposed this structure by linking DMFT, Paley--Wiener prediction theory, and Krylov operator growth, all acting on the mean-field power spectrum. The surprising point is therefore not that the chaotic network decorrelates rapidly, but that the apparent stochasticity of its mean-field description is latent determinism whose extraction has a well-defined complexity scale. This reframes the mean-field power spectrum from a passive descriptor of correlations into a diagnostic of how much temporal structure remains available in chaotic recurrent activity.

This perspective gives the theory practical content beyond the SCS model. The high-frequency decay of activity spectra can serve as an empirical probe of whether variability in structured, trained, or driven RNNs reflects irreducible fluctuations or unresolved deterministic temporal structure. The Krylov representation then turns this diagnostic into a forecasting tool by identifying which temporal modes must be resolved. This points to a possible route to memory computations in random RNNs, in which predictability is exploited by sampling readouts at multiple times rather than only at a final endpoint~\cite{Kadmon2025EfficientCoding}. More broadly, inputs, tasks, and noise are expected to reshape the deterministic and stochastic components of the mean-field spectrum. For weak external white noise, for example, the mean-field spectrum separates into deterministic and stochastic components, so the Krylov construction still bounds the reducible part of trajectory uncertainty (Sec. S10 of~\cite{SM}). That the ordering $\alpha>\lambda_L$ emerges here in a \emph{non-Hamiltonian}, dissipative system---within the bound $\lambda_L\leq 2\alpha$ proved for quantum Hamiltonian operator dynamics \cite{Parker} and conjectured classically---suggests that Krylov growth organizes chaos well beyond its original Hamiltonian setting, and leaves open why the two settings should agree.

\begin{acknowledgments}
This work was supported by the Gatsby Charitable Foundation.
\end{acknowledgments}
\nocite{DymMcKean1976,paley-weiner,SCS1988,kadmon2015,helias2020statistical,UchaikinZolotarev1999,teschl2012ordinary,teschl2000jacobi,SIMON199882,murdock2003normal,CrisantiSompolinsky2018,EngelkenWolfAbbott2023,Pazo2024,Parker,Helias2018,Walter1982}

\clearpage
\onecolumngrid

\section*{Supplemental Material}
\addcontentsline{toc}{section}{Supplemental Material}

\begin{center}
\textbf{CONTENTS}
\end{center}

\setcounter{section}{0}
\setcounter{subsection}{0}
\setcounter{subsubsection}{0}
\setcounter{equation}{0}
\setcounter{figure}{0}

\renewcommand{\thesection}{S\arabic{section}}
\renewcommand{\thesubsection}{\thesection.\Alph{subsection}}
\renewcommand{\thesubsubsection}{\arabic{subsubsection}}
\renewcommand{\theequation}{S\arabic{equation}}
\renewcommand{\thefigure}{S\arabic{figure}}

\renewcommand{\theHequation}{S\arabic{equation}}
\renewcommand{\theHfigure}{S\arabic{figure}}

\setcounter{secnumdepth}{2}

\startcontents[supp]
\printcontents[supp]{l}{1}{\setcounter{tocdepth}{2}}




\title{Supplemental Material of ``The Predictable Mean-Field Chaos in Random Recurrent Networks"}

\author{Alkesh Yadav}
\thanks{These authors contributed equally to this work.}
\affiliation{Edmond and Lily Safra Center for Brain Sciences, The Hebrew University of Jerusalem, 9190401, Israel}

\author{Vladimir Shaidurov}
\thanks{These authors contributed equally to this work.}
\affiliation{Edmond and Lily Safra Center for Brain Sciences, The Hebrew University of Jerusalem, 9190401, Israel}

\author{Jonathan Kadmon}
\affiliation{Edmond and Lily Safra Center for Brain Sciences, The Hebrew University of Jerusalem, 9190401, Israel}

\maketitle


\section{Derivation of the Time-Averaged Prediction Error}

In this section, we 
relate the prediction error, $\sigma_{\infty}^2(\tpred)$, to the power spectral density (PSD) of the Gaussian process. 
The parameter $\tpred$ serves as the horizon over which the historical state of the system provides useful information for predicting its future.

Consider a stationary zero-mean Gaussian stochastic process $X(t)$ driven by Gaussian white noise $\xi(t)$ from the infinite past ($-\infty < t < \infty$). The state of the system at time $t$ can be expressed as a convolution with the causal impulse response function $G(t)$:
\begin{equation}
    X(t) = \int_{-\infty}^t G(t-s) \xi(s) \, ds.
\end{equation}
Suppose we have observed the entire history of the process up to $t=0$, denoted as $X(s \le 0)$. The variance of the prediction error at a future time $t > 0$ is defined as:
\begin{equation}
    \sigma^2(t) := \mathbb{E}\left[ \left( X(t) - \mathbb{E}[X(t) \mid X(s \le 0)] \right)^2 \right].
\end{equation}
To evaluate this, we split the integral for $X(t)$ into the known past and the unknown future:
\begin{equation}
    X(t) = \int_{-\infty}^0 G(t-s) \xi(s) \, ds + \int_0^t G(t-s) \xi(s) \, ds.
\end{equation}
Because the noise $\xi(s)$ for $s > 0$ is independent of the past, the conditional expectation is exactly the first term. Therefore, the prediction error simplifies to:
\begin{equation}
    X(t) - \mathbb{E}[X(t) \mid X(s \le 0)] = \int_0^t G(t-s) \xi(s) \, ds.
\end{equation}
Using the It\^o isometry (or equivalently, the white noise property $\langle \xi(s)\xi(s') \rangle = \delta(s-s')$), the variance of the prediction error becomes:
\begin{equation} \label{eq:sigma_time}
    \sigma^2(t) = \int_0^t |G(t-s)|^2 \, ds = \int_0^t |G(u)|^2 \, du.
\end{equation}

We now relate the time-domain prediction error to the spectral properties of the system, leveraging a Laplace-weighted form of the Kolmogorov--Wiener prediction-error formula (following the formulation established by Dym and McKean \cite[Sec.~4.3]{DymMcKean1976}). 

Let $\tilde{\Delta}(\omega)$ be the power spectral density of the process. The Fourier transform of the response function, $\widehat{G}(\omega)$, is the causal and minimal-phase (all poles and zeros in left half plane of the complex frequency, $i\omega$) 
spectral factor of the spectral density, such that $\tilde{\Delta}(\omega) = |\widehat{G}(\omega)|^2$.
We are interested in the Laplace-weighted time average of the prediction error variance over the scale $\tpred$:
\begin{equation}
\label{eq:sigma_infty}
    \sigma^2_\infty(\tpred) =\frac{2}{\tpred} \int_0^{\infty} \sigma^2(t) e^{-2 \frac{t}{\tpred}} \, dt.
\end{equation}
Integrating by parts—and noting that $\sigma^2(0)=0$ and that the boundary term at infinity vanishes for stable systems—we obtain:
\begin{equation}
    \frac{2}{\tpred}\int_0^\infty \sigma^2(t) e^{-2\frac{t}{\tpred}} \, dt = \int_0^\infty |G(t)|^2 e^{-2\frac{t}{\tpred}} \, dt.
\end{equation}
Applying Plancherel's theorem to the function $G(t)e^{-t/\tpred}$, this integral can be expressed in the frequency domain as:
\begin{equation} \label{eq:plancherel_result}
    \sigma^2_\infty(\tpred) = \frac{1}{2\pi}\int_{-\infty}^\infty \big|\widehat{G}(\omega+i\tpred^{-1})\big|^2 \, d\omega.
\end{equation}
Because $G(t)$ is causal and represents a minimal-phase system, $\widehat{G}(z)$ is analytic and zero-free in the upper half of the complex plane ($\text{Im}(z) > 0$). Consequently, the function $\log|\widehat{G}(z)|$ is harmonic in the upper half-plane and admits a Poisson integral representation\cite[Sec. 1.9]{DymMcKean1976}. Evaluating this at $z = \omega+i\tpred^{-1}$ yields:
\begin{equation}
    \log|\widehat{G}(\omega+i\tpred^{-1})| = \frac{1}{\pi\tpred}\int_{-\infty}^\infty\frac{\log|\widehat{G}(\xi)|}{(\xi-\omega)^2+\tpred^{-2}} \, d\xi.
\end{equation}
Multiplying by $2$ and exponentiating both sides, while using the relation $\tilde \Delta(\xi) = |\widehat{G}(\xi)|^2$, we find:
\begin{equation}
    \big|\widehat{G}(\omega+i\tpred^{-1})\big|^2 = \exp\left(\frac{1}{\pi\tpred}\int_{-\infty}^\infty\frac{\log \tilde{\Delta}(\xi)}{(\xi-\omega)^2+\tpred^{-2}} \, d\xi\right).
\end{equation}
Substituting this result back into Eq.~\eqref{eq:plancherel_result} gives the final relationship:
\begin{equation} \label{eq:final_proof_result}
    \frac{2}{\tpred}\int_0^{\infty} \sigma^2(t) \, e^{-2 \frac{t}{\tpred}} \, dt = \frac{1}{2\pi}\int_{-\infty}^{\infty} \exp\left(\frac{1}{\pi\tpred}\int_{-\infty}^{\infty}\frac{\log \tilde{\Delta}(\xi)}{(\xi-\omega)^2+\tpred^{-2}} \, d\xi\right) d\omega.
\end{equation}
In the limiting case, the Paley--Wiener integral:
\begin{equation} 
\label{eq:payley_weiner}\int_{-\infty}^{\infty} \frac{\ln \tilde{\Delta}(\omega)}{1 + \omega^2} \, d\omega
\end{equation} 
diverges to \(-\infty\), and the GP corresponds to a deterministic process: the
prediction-error variance vanishes,
$\sigma_{\infty}(\tpred)=0$ for all $\tpred$.

\section{ PSD Asymptotics and Imaginary-Time Singularities of Autocorrelation}
\label{sec:fourier_tail}

In this section, we establish the connection between the high-frequency asymptotic tail of the PSD $\tilde{\Delta}(\omega)$ and the location of the nearest singularity of the autocorrelation function $\Delta(\tau)$ in the complex time plane $z = \tau + i t$.
The derivation is adapted from~\cite{paley-weiner}.
Specifically, we show that the exponential decay rate of the high-frequency tail $\omega \to \infty$ is uniquely determined by the distance $t_S$ from the real-time axis to the nearest singularity in the upper-half imaginary-time plane.
We assume that $\Delta(\tau)$ can be analytically continued into the complex time plane $z = \tau + i t$ and is holomorphic within the strip $0 \le \text{Im}(z) < t_S$, thus the boundary of this strip is determined by the nearest singularity of $\Delta(z)$ located at $z_S = it_S$. 
The endpoint convergence of the shifted contour follows from the mean-field equation. Since \(\Delta(\tau)\to\Delta_\infty\) as \(\tau\to\pm\infty\), and \(V(\Delta)\sim E-\lambda^2(\Delta-\Delta_\infty)^2/2\) near \(\Delta_\infty\), the linearized dynamics gives exponential relaxation to \(\Delta_\infty\). Along \(z=\tau+it\), with fixed \(t\), \(\Delta(\tau+it)-\Delta_\infty\sim A_+(t)e^{-\lambda\tau}\) as \(\tau\to+\infty\), and similarly with the exponent reversed as \(\tau\to-\infty\). Therefore \((\Delta(z)-\Delta_\infty)e^{i\omega z}\to0\) as \(\operatorname{Re}z\to\pm\infty\). The constant asymptotic value contributes only a zero-frequency spectral atom, corresponding to an independent static Gaussian component of the process. We therefore subtract this constant contribution and keep the notation
\(\Delta(\tau)\) for the decaying part of the autocorrelation below.

For a positive frequency $\omega > 0$, we deform the integration path upward into the complex plane along the line $\text{Im}(z) = t$, where $0 < t < t_S$. By Cauchy's integral theorem, and $\Delta(z) e^{i\omega z} \to 0$ as $\text{Re}(z) \to \pm\infty$ within the strip, the integral along the real axis equals the integral along the shifted line:
\begin{equation}
    \tilde{\Delta}(\omega) =
    \int_{-\infty}^{\infty}
    \Delta(\tau)e^{i\omega\tau}\,d\tau =\int_{-\infty}^{\infty} \Delta(\tau + it) e^{i\omega (\tau + it)} \, d\tau = e^{- \omega t} \int_{-\infty}^{\infty} \Delta(\tau + it) e^{i\omega \tau} \, d\tau.
    \label{eq:shifted_fourier}
\end{equation}
Let us define the remaining integral factor as $I(\omega, t) \equiv \int_{-\infty}^{\infty} \Delta(\tau + it) e^{i\omega \tau} \, d\tau$. Because $t<t_S$, the slice $\Delta(\tau+i t)$ lies strictly within the
domain of analyticity. The endpoint relaxation discussed above implies
$\Delta(\tau+i t)\in L^1(\mathbb R)$. By the Riemann-Lebesgue lemma, the integral vanishes at high frequencies, $\lim_{\omega \to \infty} I(\omega, t) = 0$. 
Since $I(\omega, t)$ is bounded by a finite constant $M(t) = \int_{-\infty}^{\infty} |\Delta(\tau + it)| \, d\tau$, we obtain the inequality
\begin{equation}
    |\tilde{\Delta}(\omega)| \le M(t) e^{-\omega t}.
\end{equation}
Because this inequality holds for any arbitrary choice of $t < t_S$, we can take the limit $t \to t_S^-$, yielding the asymptotic upper bound:
\begin{equation}
    \limsup_{\omega \to \infty} \frac{\ln |\tilde{\Delta}(\omega)|}{\omega} \le -t_S.
    \label{eq:upper_bound}
\end{equation}
This confirms that the Fourier transform decays at least as fast as $e^{-\omega t_S}$.

To prove that the decay cannot be strictly faster than $e^{-\omega t_S}$, we utilize the inverse Fourier transform to reconstruct the complex-time function:
\begin{equation}
    \Delta(\tau + it) = \frac{1}{2\pi} \int_{-\infty}^{\infty} \tilde{\Delta}(\omega) e^{-i\omega (\tau + it)} \, d\omega = \frac{1}{2\pi} \int_{-\infty}^{\infty} \tilde{\Delta}(\omega) e^{\omega t} e^{-i\omega \tau} \, d\omega.
\end{equation}
A sufficient condition for the inverse Fourier representation to define an
analytic continuation to imaginary-time depth $t>0$ is absolute convergence
of the weighted integral. Focusing on the $\omega > 0$ contribution, the absolute integrand is bounded by $|\tilde{\Delta}(\omega)| e^{\omega t}$. 
Suppose, for sake of contradiction, that $\tilde{\Delta}(\omega)$ decayed exponentially faster than the physical limit set by $t_S$, such that $|\tilde{\Delta}(\omega)| \le C e^{-\omega t_1}$ for some $t_1 > t_S$ as $\omega \to \infty$. Under this assumption, the integrand satisfies
\begin{equation}
    |\tilde{\Delta}(\omega)| e^{\omega t} \le C e^{-(t_1 - t)\omega}.
\end{equation}
This integral converges absolutely and uniformly for all imaginary times $t < t_1$. By the properties of the Fourier integral, the reconstructed function $\Delta(z)$ would be analytic in the extended strip $0 \le \text{Im}(z) < t_1$. 
This directly contradicts the existence of the singularity at $\text{Im}(z) = t_S < t_1$. Thus, the high-frequency tail $\tilde{\Delta}(\omega)$ cannot decay faster than $e^{-\omega t_S}$.

The combined forward and inverse arguments establish that the exponential envelope of the high-frequency Fourier tail is rigidly locked to the nearest imaginary-time singularity:
\begin{equation}
    \tilde{\Delta}(\omega) \sim P(\omega) e^{-\omega t_S} \quad \text{as} \quad \omega \to \infty,
\end{equation}
where $P(\omega)$ is a sub-exponential prefactor. The precise mathematical form of $P(\omega)$ is governed by the nature of the singularity at $z_S = it_S$.

\section{Mapping DMFT Autocorrelation Dynamics to an Effective Potential}

Following the dynamical mean-field theory (DMFT) approach for random neural networks \cite{SCS1988}\cite{kadmon2015}\cite{helias2020statistical}, the effective single-unit dynamics are governed by the Langevin-like equation
\begin{equation}
    \dot{x}(t) = - x(t) + \eta(t),
\end{equation}
where $\eta(t)$ is a zero-mean, self-consistently determined Gaussian process. In the stationary state, the statistics of $\eta(t)$ are completely specified by the network autocorrelation function, $\Delta(\tau) \equiv \langle x(t)x(t+\tau) \rangle = \langle x(0)x(\tau) \rangle$, where $\tau = |t-s|$ is the time lag. $\Delta(\tau)$ satisfies the following second-order differential equation:
\begin{equation} \label{eq:delta_eom}
    \left( -\partial_\tau^2 + 1 \right) \Delta(\tau) = g^2\langle \phi(x(0)) \phi(x(\tau)) \rangle.
\end{equation}
Because the right-hand side of Eq.~\eqref{eq:delta_eom} depends on $\tau$ exclusively through $\Delta(\tau)$, the dynamics of the autocorrelation function can be exactly mapped onto the motion of a classical particle in a conservative one-dimensional potential:
\begin{equation} \label{eq:newtonian}
    \partial_{\tau}^2 \Delta(\tau) = -\partial_\Delta V(\Delta(\tau), \Delta_0).
\end{equation}
Using standard properties of Gaussian integration, this self-consistent potential is evaluated as:
\begin{equation} \label{eq:potential}
    V(\Delta(\tau), \Delta_0) = - \frac{\Delta^2}{2} + g^2 \left( \langle \Phi(x_1) \Phi(x_2) \rangle_{\mathbf{x} \sim \mathcal{N}(0,\Sigma)} - \langle \Phi(x)\rangle_{x\sim \mathcal{N}(0,\Delta_0)}^2 \right),
\end{equation}
where $\Phi(x) = \int \phi(x) \, dx$, and the covariance matrix for the bivariate normal distribution is given by
\begin{equation}
    \Sigma = 
    \begin{bmatrix}
        \Delta_0 & \Delta(\tau) \\
        \Delta(\tau) & \Delta_0
    \end{bmatrix}.
\end{equation}
The stationary autocorrelation must be an even function, $\Delta(\tau) = \Delta(-\tau)$, which is gauranteed by
the dynamics of $\Delta(\tau)$ being time-reversal invariant---if $\Delta(\tau)$ is a solution, so is $\Delta(-\tau)$. 
The potential $V$ is at least once-differentiable, the physical trajectory cannot develop a non-physical cusp at the origin. 
Together, these mathematical constraints enforce the initial condition $\dot{\Delta}(0) = 0$.

We can then use this condition to self-consistently determine the initial amplitude, $\Delta_0 \equiv \Delta(0)$. 
The energy conservation of the classical particle yields: $\frac{1}{2}\dot{\Delta}(\tau)^2 + V(\Delta(\tau), \Delta_0) = E$. Assuming the network correlations decay to zero at large time lags ($\Delta(\infty) = \dot{\Delta}(\infty) = 0$), and noting from Eq.~\eqref{eq:potential} that $V(0, \Delta_0) = 0$, the total energy is $E = 0$.  $\dot{\Delta}(0) = 0$ directly yields the self-consistency condition for the initial amplitude:
\( 
    V(\Delta_0, \Delta_0) = 0.
\)

Once $\Delta_0$ is determined, the potential $V(\Delta) \equiv V(\Delta, \Delta_0)$ is fully parameterized, and the problem reduces to a simple classical mechanics problem of a particle escaping a fixed potential well. $\Delta(\tau)$ is given by the autonomous initial value problem:
\begin{equation} \label{eq:ivp}
    \partial_\tau^2 \Delta(\tau) = - \partial_\Delta V(\Delta), \quad \text{with} \quad \Delta(0) = \Delta_0, \quad \dot{\Delta}(0) = 0.
\end{equation}
The regularity of the global trajectory is entirely controlled by the regularity of the effective force term, $-\partial_\Delta V$.

\section{\texorpdfstring{Analytic Structure of the Potential, $V(\Delta)$}{Analytic Structure of the Potential, V(Delta)}}
\label{sec:potential_analyticity}
Since the first term in the potential trivially continues to the complex plane, we focus on the expectation term which decides the analytic nature of the potential, $V(\Delta)$.
\begin{equation}
    E(\Delta) = \langle \Phi(\sqrt{\Delta_0} x_1) \Phi(\sqrt{\Delta_0}x_2) \rangle_{x_1,x_2 \sim \mathcal{N}(0,\Sigma_{\rm std})}; \quad \Sigma_{\rm std} = 
    \begin{bmatrix}
        1 & \Delta/\Delta_0 \\
        \Delta/\Delta_0 & 1
    \end{bmatrix}
\end{equation}
To understand this analytic structure of $E(\Delta)$ in the complex $\Delta$-plane, we use the Hermite polynomial expansion. This expansion can be performed for any square integrable non-linearity with respect to the Gaussian measure. Let $H_n(x) =  (-1)^n e^{x^2/2} \partial_x^n e^{-x^2/2}$, be the probabilist's Hermite polynomials given by the Rodrigues formula. We can expand $\Phi(x)$ as:
\begin{equation}
    \Phi(x) = \sum_{n=0}^\infty \frac{A_n}{n!} H_n(x) \quad \text{where} \quad A_n =  \langle \Phi(\sqrt{\Delta_0}x) H_n(x) \rangle_{x \sim \mathcal{N}(0,1)} 
\end{equation} 
Because $x_1$ and $x_2$ are jointly Gaussian with correlation $\Delta/\Delta_0$, the orthogonality of Hermite polynomials gives a Taylor series for $E(\Delta)$ around $\Delta = 0$:
\begin{equation}
    \label{eq:E_delta}
    E(\Delta) = \sum_{n=0}^\infty \sum_{n'=0}^\infty \frac{A_n}{n!}\frac{A_{n'}}{n'!}\langle H_n(x_1) H_{n'}(x_2) \rangle_{x_1,x_2 \sim \mathcal{N}(0,\Sigma_{\rm std}) } = \sum_{n=0}^\infty \frac{A_n^2}{n!\Delta_0^n} \Delta^n 
\end{equation}
where we have used $\langle H_n(x_1) H_{n'}(x_2) \rangle_{x_1,x_2 \sim \mathcal{N}(0,\Sigma_{\rm std})} = n! (\Delta/\Delta_0)^n \delta_{n,n'}$. 
The analytic structure of $E(\Delta)$, and hence $V(\Delta)$, is completely dictated by the radius of convergence of this series and its analytic continuation. We express the coefficients, $A_n$, using the Fourier transform $\hat{\Phi}(k) = \int \Phi(x) e^{-ikx} dx$:
\begin{equation} 
\begin{aligned}
    A_n & =  \langle \Phi(\sqrt{\Delta_0}x) H_n(x) \rangle_{x \sim \mathcal{N}(0,1)} = \frac{(-1)^n}{\sqrt{2\pi}}\int_{-\infty}^\infty \Phi(\sqrt{\Delta_0}x) \partial_x^n e^{-x^2/2} dx = \langle \partial_x^n\Phi(\sqrt{\Delta_0}x) \rangle_{x \sim \mathcal{N}(0,1)} \\ 
     & = \frac{i^n \Delta_0^{n/2}}{2\pi} \int_{-\infty}^\infty \hat{\Phi}(k) \, k^n e^{-\Delta_0k^2/2} dk
\end{aligned}
\end{equation}

Decompose the Fourier transform as $\widehat{\Phi}(k)=\widehat{\Phi}_e(k)+i\widehat{\Phi}_o(k)$, where $\widehat{\Phi}_e$ is even and $\widehat{\Phi}_o$ is odd. By parity, $A_{2n}=\frac{(-1)^n \Delta_0^{n}}{2\pi} \int_{-\infty}^\infty \widehat{\Phi}_e(k) \, k^{2n} e^{-\Delta_0k^2/2} dk$
	and $A_{2n+1}=\frac{(-1)^{n+1} \Delta_0^{n+\frac{1}{2}}}{2\pi} \int_{-\infty}^\infty \widehat{\Phi}_o(k) \, k^{2n+1} e^{-\Delta_0k^2/2} dk$.
Let the Fourier tail of the nonlinearity be given by $\hat{\Phi}(k) \sim \exp(- |k|^\beta/k_0)$. The global Fourier tail determines the slower decay between the two parity parts. The other parity sector decays at least as fast, and may decay faster.
For definiteness, we assume that the even part is the slowest-decaying one;
the case in which the odd sector is slowest is obtained by the identical
argument applied to the odd subsequence. The saddle analysis is then applied to each parity subsequence separately.
The full radius of convergence is controlled by the slowest-decaying subsequence. When the corresponding radii of convergence are different, this subsequence also determines the nearest singularity
type. If the radii are the same, then adding their local singular expansions at the relevant boundary point shows that the singularity cannot be canceled, since all coefficients are positive.

\begin{equation}
	\label{eq:saddle_point_coeff}
	A_{2n}
	=
	\frac{(-1)^n\Delta_0^{n}}{2\pi}
	\int_{-\infty}^\infty
	\widehat{\Phi}(k)\, k^{2n} e^{-\Delta_0k^2/2}\, dk
	\approx
	\frac{(-1)^n \Delta_0^{n}}{\pi}
	\sqrt{\frac{2\pi}{S_n''(k_s)}} e^{-S_n(k_s)} .
\end{equation}
Here the \(n\)-dependent action is \(S_n(k)=-2n\ln |k|+|k|^\beta/k_0+\Delta_0 k^2/2\), and the positive saddle \(k_s=k_s(n)\) is determined by \(\Delta_0 k_s^2+\beta k_s^\beta/k_0=2n\), with \(k_s>0\). Using the saddle-point equation, we eliminate the term \(\Delta_0 k_s^2/2\) in the action:
\[
S_n(k_s)
=
-n\ln(k_s^2)
+
n
+
\frac{2-\beta}{2k_0} k_s^\beta .
\]

We define \(x_n:=\beta k_s^\beta/(2n k_0)\), which characterizes the relative contribution of the Fourier-tail term to the saddle equation. The saddle-point equation then becomes
\(
\Delta_0 k_s^2 = 2n(1-x_n).
\)
Since both terms in the original saddle-point equation are positive, \(0<x_n<1\). Raising this equation to the power \(\beta/2\) and substituting into the definition of \(x_n\), we obtain
\begin{equation}
\label{eq:saddle_point}
	x_n
	=
	\frac{\beta}{k_0\Delta_0^{\beta/2}}
	(2n)^{\beta/2-1}
	(1-x_n)^{\beta/2}=C_n(1-x_n)^{\beta/2}
\end{equation}
where \(C_n:=\beta(2n)^{\beta/2-1}/(k_0\Delta_0^{\beta/2})\).
Finally, substituting \(k_s^2=2n(1-x_n)/\Delta_0\) and \(k_s^\beta/k_0=2n x_n/\beta\) into the action, we obtain
\begin{equation} 
	\label{eq:action}
	S_n(x_n)
	=
	-n\ln\frac{2n}{\Delta_0}
	+
	n
	-
	n\ln(1-x_n)
	+
	n\left(\frac{2}{\beta}-1\right)x_n .
\end{equation} 

From the saddle approximation in Eq.~\eqref{eq:saddle_point}, we have
\[
\log\left(\frac{|A_{2n}|^2}{(2n)!}\right)
\approx
2n\log\Delta_0
-2S_n(x_n)
-\log((2n)!)-\log(S_n''(x_n))
\]
Since $
S_n''(k_s)
=
2\Delta_0
+
\frac{\beta^2}{k_0}k_s^{\beta-2}
=
\Delta_0\left(
2+\beta\frac{x_n}{1-x_n}
\right)
=
\Delta_0
\frac{2+(\beta-2)x_n}{1-x_n}
$,  we have
\[
\log S_n''(k_s)
=
\log\Delta_0
+
\log\bigl(2+(\beta-2)x_n\bigr)
-
\log(1-x_n).
\]
Using Stirling's formula, \(\log((2n)!)=2n\log(2n)-2n+\frac{1}{2}\log(4\pi n)+O(1)\), and substituting Eq.~\eqref{eq:action}, we obtain
\[
\begin{aligned}
	\log\left(\frac{|A_{2n}|^2}{(2n)!}\right)
	&=
	2n\log\Delta_0
	+
	2n\log\frac{2n}{\Delta_0}
	-2n
	+2n\log(1-x_n)
	-2n\left(\frac{2}{\beta}-1\right)x_n  \\
	&\hspace{1cm}
	-2n\log(2n)+2n-\frac{1}{2}\log(4\pi n)
	+\log(1-x_n)
	-\log\Delta_0
	-\log\bigl(2+(\beta-2)x_n\bigr)
\end{aligned}
\]
After cancellation of the leading Stirling terms, this becomes
\begin{equation}
	\label{eq:log_fact}
	\log\left(\frac{|A_{2n}|^2}{(2n)!}\right)
	\approx 
	2n\left[
	\log(1-x_n)
	+
	\left(1-\frac{2}{\beta}\right)x_n
	\right]+ \log(1-x_n)-\log\bigl(2+(\beta-2)x_n\bigr)-\frac{1}{2}\log(4\pi n)
\end{equation}

This is the key form for the rest of the argument: the large-\(n\) behavior is
controlled by \(x_n\), which is determined implicitly by
Eq.~\eqref{eq:saddle_point}. Since \(C_n\to0\) for \(\beta<2\), is constant
for \(\beta=2\), and \(C_n\to\infty\) for \(\beta>2\), the three regimes must
be treated separately. We first use Eq.~\eqref{eq:log_fact} to determine the
radius of convergence from the Cauchy--Hadamard formula,
\begin{equation}
	\label{eq:CH}
	\frac{1}{R}
	=
	\limsup_{n\to\infty}
	\left(
	\frac{|A_n|^2}{n!\Delta_0^n}
	\right)^{1/n}
	=
	\frac{1}{\Delta_0}
	\limsup_{n\to\infty}
	\exp\left[
	\frac{1}{n}\log\left(\frac{|A_n|^2}{n!}\right)
	\right].
\end{equation}
 Once the radius is fixed, the same coefficient
asymptotics suggest the scale and type of the boundary singularity. The asymptotic series alone does not control beyond-all-orders terms, which may be
exponentially small in one direction of approach but become dominant after
analytic continuation to another. A rigorous proof would therefore require
uniform control of the analytically continued remainder in all directions of
approach to the singular point. Nethertheless, for
concrete nonlinearities such as \(\phi(x)=\tanh x\), and later for
Gaussian-smoothed piecewise-polynomial functions, we check the predicted
singularity directly by contour methods.

\subsection{\texorpdfstring{Smooth activation functions with $0<\beta<2$}{Smooth activation functions with 0 < beta < 2}}
\subsubsection*{Analysis of the radius of convergence}
When \(0<\beta<2\), the prefactor \(C_n\) in Eq \ref{eq:saddle_point} to zero as \(n\to\infty\). Since \(0<x_n<1\), it follows that \((1-x_n)^{\beta/2}<1\), and therefore \(x_n<C_n\). Hence \(x_n\to0\). Expanding \(x_n=C_n(1-x_n)^{\beta/2}\) for small \(x_n\) and solving iteratively, we obtain
\[
x_n
=
C_n\left(1-\frac{\beta}{2}x_n+O(x_n^2)\right)
=
C_n-\frac{\beta}{2}C_nx_n+O(C_n^3).
\]
since \(x_n=O(C_n)\). Therefore we get
\(
\log(1-x_n)
+
\left(1-\frac{2}{\beta}\right)x_n
=
-\frac{2}{\beta}C_n
+
\frac{1}{2}C_n^2
+
O(C_n^3).
\)
Substituting this into Eq.~\eqref{eq:log_fact},
\begin{align}
\log\left(\frac{|A_{2n}|^2}{(2n)!}\right)
& \approx
2n\left(
-\frac{2}{\beta}C_n+\frac{1}{2}C_n^2
\right)-\log 2-\frac{\beta}{2}C_n \nonumber\\
 & = 2n\left(
-\frac{2}{k_0\Delta_0^{\beta/2}}(2n)^{\beta/2-1}
+
\frac{\beta^2}{2k_0^2\Delta_0^\beta}(2n)^{\beta-2}
-\frac{\log 2}{2n}
-\frac{\beta^2}{2k_0\Delta_0^{\beta/2}}(2n)^{\beta/2-2}
\right).
\end{align}

Dividing by \(2n\), we see that all terms vanish for \(0<\beta<2\). Hence
\[
\lim_{2n\to\infty}
\frac{1}{2n}
\log\left(\frac{|A_{2n}|^2}{(2n)!}\right)
=0.
\]
Using Eq.~\eqref{eq:CH}, we obtain \(1/R=1/\Delta_0\), and therefore \(R=\Delta_0\).

For \(\phi(x)=\tanh x\), we have \(\Phi(x)=\log\cosh x\). The nearest poles of \(\phi\) are at \(z=\pm i\pi/2\), so the Paley--Wiener estimate gives \(\widehat{\phi}(k)\sim e^{-\pi|k|/2}\), and therefore \(\widehat{\Phi}(k)\sim k^{-1}e^{-\pi|k|/2}\). This corresponds to \(\beta=1<2\), so the radius of convergence is again \(R=\Delta_0\).
\subsubsection*{Analysis of the singularity}
\paragraph{Singularity predicted by the coefficient asymptotics.} The coefficient asymptotics have the leading form
\begin{equation}
	\frac{A_{2n}^2}{(2n)!}
	\doteq A \exp(-a n^{\beta/2}),
\end{equation}
for constants \(A,a>0\), where \(\doteq\) denotes leading-order equality as
\(n\to\infty\). Since \(0<\beta/2<1\), the stretched exponential admits the
Laplace representation
\begin{equation}
	e^{-a n^{\beta/2}} = \int_0^\infty e^{-nt} g(t) \, dt.
\end{equation}
Here, \(g\) is the density of a positive \(\beta/2\)-stable Lévy distribution.
Defining the dimensionless parameter $\rho = \Delta/\Delta_0$, we can express the sum $E(\rho)$ by exchanging the sum and the integral:
\begin{equation}
	\begin{aligned}
		E(\rho) = \sum_{n=0}^\infty A \rho^{2n} \left( \int_0^\infty e^{-nt} g(t) \, dt \right) 
		= A \int_0^\infty \frac{g(t)}{1 - \rho^2 e^{-t}} \, dt.
	\end{aligned}
	\label{eq:E_integral_general}
\end{equation}
Although $g(t)$ lacks an elementary closed form for arbitrary $\beta$, Eq.~\eqref{eq:E_integral_general} provides a powerful integral representation that facilitates both analytic continuation and analysis of the singularity structure.

To investigate the exact case for $\beta = 1$, we utilize the known closed form of the Lévy distribution, $g(t) = \frac{a}{2\sqrt{\pi} t^{3/2}} \exp\left(-\frac{a^2}{4t}\right)$. Substituting this into Eq.~\eqref{eq:E_integral_general} yields:
\begin{equation}
	E(\rho) = A' \int_0^\infty \frac{\exp\left(-\frac{a^2}{4t}\right)}{t^{3/2}(1 - \rho^2 e^{-t})} \, dt.
	\label{eq:E_rho_exact}
\end{equation}
The integrand in Eq.~\eqref{eq:E_rho_exact} possesses two distinct types of singularities in the complex $t$-plane. First, there is a fixed singularity at the origin ($t=0$) comprising an essential singularity driven by the $\exp(-a^2/4t)$ term and a branch point from the $t^{-3/2}$ factor. Second, the integrand exhibits a moving simple pole at $t_0$, located where the denominator vanishes:
\begin{equation}
	1 - \rho^2 e^{-t} = 0 \implies t_0 = 2\ln|\rho|.
\end{equation}
The dynamics of this pole as a function of $\rho$ dictate the domain of convergence for the integral along the positive real axis:
\begin{itemize}
	\item \textbf{$|\rho| < 1$:} The pole lies on the negative real axis ($t_0 < 0$). The integrand is analytic along the entire path of integration $(0, \infty)$.
	\item \textbf{$|\rho| > 1$:} The pole moves onto the positive real axis ($t_0 > 0$), directly intersecting the integration contour. Consequently, the integral must be interpreted via analytic continuation, either by taking the Cauchy principal value or by deforming the contour into the complex plane.
\end{itemize}
Despite the presence of the essential singularity and the collision of the pole at $t=0$, the integral remains absolutely convergent at its lower limit for all $\rho \in [-1, 1]$. As $t \to 0^+$, the algebraic divergence of $t^{-3/2}$ (or $t^{-5/2}$ at $|\rho|=1$ due to the Taylor expansion of the denominator $1-e^{-t} \sim t$) is overwhelmingly suppressed by the essential singularity. Formally, for any $k > 0$:
\begin{equation}
	\lim_{t \to 0^+} \frac{\exp\left(-\frac{a^2}{4t}\right)}{t^{k}} = 0.
\end{equation}
Thus, the physical behavior on the real axis as $\rho \to \pm 1$ from within the interval $(-1, 1)$ is perfectly smooth. This is corroborated by examining the $k$-th derivative of the series with respect to $\rho$:
\begin{equation}
	E^{(k)}(\pm 1) \propto \sum_{n=k}^\infty n(n-1)\cdots(n-k+1) e^{-a\sqrt{n}}.
\end{equation}
Because the stretched exponential decays faster than any polynomial $n^k$, this sum converges for all integers $k$. We conclude that $E(\rho)$ is a $C^\infty$ (infinitely differentiable) function on the closed interval $[-1, 1]$.

The true nature of the singularity emerges only when crossing the boundary into the regime $|\rho| > 1$. Because the pole $t_0 = 2\ln|\rho| > 0$ now lies on the integration contour, the standard Lebesgue integral is undefined. To analytically continue the function, we must deform the contour in the complex $t$-plane to avoid the pole. Depending on the direction of the deformation ($\pm i\epsilon$), we pick up a branch cut jump proportional to $2\pi i$ times the residue of the integrand at $t_0$:
\begin{equation}
	\text{Res}_{t=t_0} \left[ \frac{\exp\left(-\frac{a^2}{4t}\right)}{t^{3/2}(1 - \rho^2 e^{-t})} \right] = \frac{\exp\left(-\frac{a^2}{4t_0}\right)}{t_0^{3/2} \left[ \frac{d}{dt}(1-\rho^2 e^{-t}) \right]_{t=t_0}}.
\end{equation}
Noting that $\rho^2 e^{-t_0} = 1$, the derivative evaluates to $1$. Substituting $t_0 = 2\ln|\rho|$ yields:
\begin{equation}
	\text{Res}_{t=t_0} = \frac{\exp\left(-\frac{a^2}{8\ln|\rho|}\right)}{(2\ln|\rho|)^{3/2}}.
\end{equation}
To analyze the critical behavior, let $\rho = 1 + \delta$ for $0 < \delta \ll 1$. Expanding the logarithm as $\ln|\rho| \approx \delta$, the non-analytic jump, $\delta E(\rho)$, behaves asymptotically as:
\begin{equation}
	\delta E(\rho) \propto \delta^{-3/2} \exp\left(-\frac{a^2}{8\delta}\right).
\end{equation}
Because the non-analytic behavior is entirely dominated by the factor $\exp(-\text{const}/(\rho-1))$, the point $\rho = \pm 1$ represents an essential singularity. For the original covariance, this should be understood as the singularity
suggested by the asymptotic expansion. 
For general $0<\beta<2$ a closed form expression for $g(t)$ is not available, the short time asymptotics are given by \cite[Thm.~5.4.1]{UchaikinZolotarev1999}: 
\begin{equation}
    g(t) \sim \exp(-C t^{-\beta/(2-\beta)})
\end{equation}
Therefore,the derivative $E(\rho)$ is asymptotically given by 
\begin{equation}
\label{eq:assymptotics_beta}
    E'(\rho) \sim  \int_0^\infty \frac{\exp(-C t^{-\beta/(2-\beta)})}{(1 - \rho^2 e^{-t})} \, dt.
\end{equation}
With the same calculation as in the $\beta=1$ case, $E(\rho)$ is analytical in the region $\rho\in (-1,1)$, but for $|\rho|>1$, $E(\rho)$ develops a essential singularity with of the form $\delta E(\rho) \sim \exp({-C\delta^{-\frac{\beta}{2-\beta}}})$. 
We now do a non-asymptotic calculation to analyze the specific non-linearity $\phi(x) = \tanh(x)$, and show that it agrees with the asymptotic calculation.

\paragraph{Singularity of the potential for $\phi(x)=\tanh x$}
Here we show, for $\phi(x)=\tanh x$, that
		\[
		E(1+\delta\pm i0)\approx B_{\pm}\,\delta^{-1/2}
		\exp\left(-\frac{\pi^2}{4\delta}\right)
		\]
Recall that
	\[
	E'(\rho)
	=
	\frac{1}{2\pi\sqrt{1-\rho^2}}
	\int_{\mathbb R^2}
	\tanh x\,\tanh y\,
	\exp\left(
	-\frac{x^2-2\rho xy+y^2}{2(1-\rho^2)}
	\right)
	\,dx\,dy .
	\]
	Write \(\rho=1-\varepsilon\), \(x=\frac{u+v}{\sqrt{2}}\), and \(y=\frac{u-v}{\sqrt{2}}\). With
		\((1-\rho^2)=\varepsilon(2-\varepsilon)\),
	for \(\operatorname{Re}\varepsilon>0\), we get

	\begin{equation}
	E'(1-\varepsilon)
	=
	\frac{1}{2\pi\sqrt{\varepsilon(2-\varepsilon)}}
	\int_{\mathbb R^2}
	\tanh\left(\frac{u+v}{\sqrt2}\right)
	\tanh\left(\frac{u-v}{\sqrt2}\right)
	\exp\left[
	-\frac{u^2}{2(2-\varepsilon)}
	-\frac{v^2}{2\varepsilon}
	\right]\,du\,dv,
		\end{equation}
		For fixed \(u\), we isolate the one-dimensional contour integral
	\begin{equation}
	I(u,\varepsilon)
	:=
	\int_C h(v)\exp\left(-\frac{v^2}{2\varepsilon}\right)\,dv .
	\end{equation}
		This integral diverges once $\operatorname{Re}\varepsilon<0$. We define its analytic continuation for arbitrary $\varepsilon$ using a steepest-descent contour.
	If \(\varepsilon=|\varepsilon|e^{i\theta}\) and \(v=re^{i\phi}\), then the
	steepest-descent directions for the phase \(-v^2/(2\varepsilon)\) satisfy
	\[
	2\phi-\theta\equiv 0 \pmod{2\pi},
	\qquad\text{so}\qquad
	\phi=\frac{\theta}{2}\pmod{\pi} .
	\]
	We take the branch cut in the \(\varepsilon\)-plane along
	\(
	\varepsilon\in(-\infty,0).
	\)
	Since \(\rho=1-\varepsilon\), this becomes the cut
	\(
	\rho\in(1,\infty)
	\)
	in the \(\rho\)-plane. Writing \(\varepsilon=|\varepsilon|e^{i\theta}\), the two
	lateral limits on the cut correspond to
	\(
	\theta\to \pi^{-}
	\) and \(
	\theta\to -\pi^{+}.
	\)
	The steepest-descent directions for the Gaussian factor
	\(\exp[-v^2/(2\varepsilon)]\) are given by
	\(
	\phi=\frac{\theta}{2}\pmod{\pi}.
	\)
	Hence
	\[
	\theta\to \pi^{-}
	\quad\Rightarrow\quad
	\phi\to \frac{\pi}{2}^{-},
	\qquad
	\theta\to -\pi^{+}
	\quad\Rightarrow\quad
	\phi\to -\frac{\pi}{2}^{+}.
	\]
	Thus, when \(\varepsilon\) crosses the negative real axis, the steepest-descent
	contour changes from one oriented vertical contour to the oppositely oriented
	one.
		We now package the \(u\)-integration into
	\[
	H_\delta(s)
	:=
	\int_{\mathbb R}
	\exp\left(-\frac{u^2}{2(2+\delta)}\right)
	g(u,is)\,du .
	\]
	After rotating the \(v\)-contour to the corresponding steepest-descent
	contour and writing \(v=is\), the factor \(dv=i\,ds\) cancels the phase of
	\(\sqrt{\varepsilon}\). Thus
	\begin{equation}
	E'(1+\delta\pm i0)
	=
	\frac{1}{2\pi\sqrt{\delta(2+\delta)}}
	\int_{\mathcal C_\pm}
	\exp\left(-\frac{s^2}{2\delta}\right)
	H_\delta(s)\,ds .
	\end{equation}
	Here \(\mathcal C_\pm\) are the two lateral continuations of the real
	\(s\)-contour obtained from the two continuations
		\(\varepsilon=-\delta\mp i0\). The \(u\)-integral converges at infinity, but it becomes singular when poles
	of \( g(u,is)\) hit the real \(u\)-contour. Since the poles of
	\(\tanh z\) are \(z=i\pi(n+1/2)\), the two factors have poles at
	\[
	u=i(s_n-s),
	\qquad
	u=i(s_n+s),
	\qquad
	s_n=\frac{(2n+1)\pi}{\sqrt2}.
	\]
	For real \(s\), these poles reach the real \(u\)-contour at \(s=s_n\).
	Therefore \(H_\delta(s)\) has singularities at
	\(
	s_n=(2n+1)\pi/\sqrt2.
	\)
	The two lateral contours \(\mathcal C_\pm\) differ by how they bypass the
	singularities \(s=s_n\).
		Near these points, the corresponding boundary value of \(H_\delta\) has the
	local form
	\[
	H_\delta(s)
	\approx
	\frac{R_\pm(\delta)}{s\mp s_n},
	\qquad
		s\to \pm s_n.
		\]
	The two lateral contours \(\mathcal C_\pm\) bypass these singularities in
	opposite ways. More precisely, one of the contours passes the positive
	singularities \(s_n>0\) from above and the negative singularities \(s_n<0\)
	from below, while the other contour passes them from the opposite sides.
	Thus the two boundary values \(V(1+\delta\pm i0)\) differ by the residue
	contributions picked up at these singularities.
	\[
	\mathcal C_+:
	\begin{cases}
		\text{passes } s_n>0 \text{ from above},\\
		\text{passes } s_n<0 \text{ from below},
	\end{cases}
	\qquad
	\mathcal C_-:
	\begin{cases}
		\text{passes } s_n>0 \text{ from below},\\
		\text{passes } s_n<0 \text{ from above}.
	\end{cases}
	\]
	\[
	\eta_n^+
	=
	-\operatorname{sgn}(s_n),
	\qquad
	\eta_n^-
	=
	\operatorname{sgn}(s_n).
	\]
	Let
	\[
	R_n(\delta):=\operatorname*{Res}_{s=s_n}H_\delta(s),
	\qquad
	s_n=\frac{(2n+1)\pi}{\sqrt2}.
	\]
		For $\tanh$, $R_n(\delta)=-2\pi \operatorname{sgn}(2n+1)$.
	Then the lateral contribution of the singularities is
	\[
	\int_{\mathcal C_\pm}
	e^{-s^2/(2\delta)}H_\delta(s)\,ds
	=
	\int_{\rm reg}
	e^{-s^2/(2\delta)}H_\delta(s)\,ds
	+
	i\pi
	\sum_{n\in\mathbb Z}
	\eta_n^\pm R_n(\delta)
	\exp\left(-\frac{s_n^2}{2\delta}\right).
	\]
	\begin{equation}
	\sum_{n\in\mathbb Z}
	\eta_n^\pm R_n(\delta)
	\exp\left(-\frac{s_n^2}{2\delta}\right)
	=
	\begin{cases}
		\displaystyle
		4\pi
		\sum_{n=0}^{\infty}
		\exp\left(-\frac{(s_n^+)^2}{2\delta}\right),
		& \mathcal C_+,
		\\[1.2em]
		\displaystyle
		-4\pi
		\sum_{n=0}^{\infty}
		\exp\left(-\frac{(s_n^+)^2}{2\delta}\right),
		& \mathcal C_- .
	\end{cases}
	\end{equation}
	\begin{equation}
	E'(1+\delta\pm i0)
	=
	E'_{\rm reg}(\delta)
	\pm
	\frac{2\pi i}{\sqrt{\delta(2+\delta)}}
	\sum_{n=0}^{\infty}
	\exp\left(
	-\frac{(2n+1)^2\pi^2}{4\delta}
	\right).
	\end{equation}

\subsection{Smooth activation functions with $\beta=2$}
\subsubsection*{Analysis of the radius of convergence}
For \(\beta=2\), Eq.~\eqref{eq:saddle_point} gives
\(
1-x_n=\frac{\Delta_0}{\Delta_0+2/k_0}.
\)
In this case the term proportional to \(1-2/\beta\) in
Eq.~\eqref{eq:log_fact} vanishes, and the remaining logarithmic terms are only
\(O(1)\) in \(n\). Therefore,
\[
\lim_{n\to\infty}
\frac{1}{2n}
\log\left(\frac{|A_{2n}|^2}{(2n)!}\right)
=
\ln(1-x_n).
\]
Using Eq.~\eqref{eq:CH}, we obtain
\begin{equation}
\frac{1}{R}
=
\frac{1}{\Delta_0}\exp(\ln(1-x_n))
=
\frac{1-x_n}{\Delta_0}
=
\frac{1}{\Delta_0+2/k_0}.
\end{equation}
Thus \(R=\Delta_0+\frac{2}{k_0}\).
For $\phi(x) = \operatorname{Erf}(\frac{x}{\sqrt{2}\sigma})$, $\Phi(x) = x \operatorname{Erf}(\frac{x}{\sqrt{2}\sigma}) + \sqrt{\frac{2}{\pi}}\sigma\exp(-\frac{x^2}{2\sigma^2})$.
At high frequencies, its Fourier transform scales as $\hat{\Phi}(k) \sim \frac{1}{k^2} e^{-k^2\sigma^2/2}$, and $k_0 =\frac{2}{\sigma^2}$. Therefore, the radius of convergence is $R = \Delta_0+\sigma^2$.
\subsubsection*{Singularity analysis}
\paragraph{Singularity predicted by the coefficient asymptotics} In this case, even though the Gaussian exponential tail determines the location of the
singularity, the algebraic prefactor determines its local branch
exponent. Assume
\[
\widehat{\Phi}(k)\sim Ck^a e^{-k^2/k_0}.
\]
Then the action in Eq.~\eqref{eq:saddle_point} acquires the correction $-a\log(k)$, or more explicitly, \(S_n(k)=-(2n+a)\ln |k|+|k|^2/k_0+\Delta_0 k^2/2\).
The saddle acquires only a subleading correction:
\(
k_s^2
=
\frac{2n}{\Delta_0+\frac{2}{k_0}}
\left(1+\frac{a}{2n}\right).
\)
The curvature evaluated at the saddle is unchanged: $S''(x_n)=\frac{2\Delta_0}{1-x_n}$.
Thus Eq.~\eqref{eq:log_fact} is modified to
\[2n\log(1-x_n)+\log(1-x_n)-\log(2)+a\log(2n)-\frac{1}{2}\log(4\pi n)=2n\log(1-x_n)+(a-\frac{1}{2})\log(n)+O(1).\]
Substituting $1-x_n=\frac{\Delta_0}{\Delta_0+2/k_0}$, we get
\(
\frac{|A_{2n}|^2}{(2n)!\Delta_0^{2n}}
\sim
K\left(\frac{1}{\Delta_0+\frac{2}{k_0}}\right)^{2n}
n^{a-\frac12}.
\)
These coefficients have the same large-\(n\) form as those of the binomial
singularity
\[
\left(1-\left(\frac{\Delta}{\Delta_0+\frac{2}{k_0}}\right)^2\right)^{-\lambda}
=
\sum_{n\ge0}
(\Delta_0+\frac{2}{k_0})^{-2n}
\frac{\Gamma(n+\lambda)}{\Gamma(\lambda)\Gamma(n+1)}
\Delta^{2n},
\qquad
\frac{\Gamma(n+\lambda)}{\Gamma(n+1)}\sim n^{\lambda-1},
\]
with \(\lambda=a+\frac12\).
Hence the singular part of the potential has the local form
\begin{equation}
\label{eq:beta_2_sing_type}
V_{\rm sing}(\Delta)
\sim
K \left(
1-\frac{\Delta}{R}
\right)^{-\left(a+\frac12\right)}.
\end{equation}

\paragraph{Gaussian-smoothed piecewise-polynomials} For a Gaussian-smoothed piecewise-polynomial function, the leading Fourier tail
is more generally a finite oscillatory sum,
\begin{equation}
\widehat{\Phi}(k)
=
e^{-k^2/k_0}
\sum_j c_j k^{a_j}e^{-ikx_j},
\end{equation}
where the points \(x_j\) are the breakpoints and the powers \(a_j\) are
determined by the first nonzero jump in the derivatives at \(x_j\). Thus
\(E\) is a sum over pairs of breakpoints. After setting
\(u=(k+l)/\sqrt2\), \(v=(k-l)/\sqrt2\), we obtain
\begin{equation}
E'(\Delta)
=
K\sum_{i,j}c_i c_j
\iint
\left(\frac{u+v}{\sqrt2}\right)^{a_i}
\left(\frac{u-v}{\sqrt2}\right)^{a_j}
e^{-i\frac{x_i+x_j}{\sqrt2}u}
e^{-i\frac{x_i-x_j}{\sqrt2}v}
e^{-\frac{\Delta_0+2/k_0+\Delta}{2}u^2
	-\frac{\Delta_0+2/k_0-\Delta}{2}v^2}
\,du\,dv .
\end{equation}
Introduce
\(
\epsilon=\Delta_0+\frac{2}{k_0}-\Delta .
\)
The \(u\)-direction remains Gaussian-damped, while the possible singularity
comes from the large-\(v\) region. Defining
\begin{equation}
H^{(i,j)}_\epsilon(v)
=
\int
e^{-\frac{2(\Delta_0+2/k_0)-\epsilon}{2}u^2}
e^{-i\frac{x_i+x_j}{\sqrt2}u}
\left(\frac{u+v}{\sqrt2}\right)^{a_i}
\left(\frac{u-v}{\sqrt2}\right)^{a_j}
\,du ,
\end{equation}
we have \(H^{(i,j)}_\epsilon(v)\sim K_{ij}v^{a_i+a_j}\) as
\(|v|\to\infty\). Thus each pair of breakpoints contributes, at the singular
scale,
\begin{equation}
E_{ij}^{\rm sing}
\sim
K_{ij}
\int^\infty v^{a_i+a_j}e^{-\epsilon v^2/2}e^{-id_{ij}v}\,dv,
\qquad
d_{ij}=\frac{x_i-x_j}{\sqrt2}.
\end{equation}
If \(d_{ij}\ne0\), steepest descent gives
\begin{equation}
E_{ij}^{\rm sing}
\sim
K_{ij}(-id_{ij})^{a_i+a_j}
\epsilon^{-a_i-a_j-\frac12}
\exp\left[-\frac{d_{ij}^2}{2\epsilon}\right],
\end{equation}
so these off-diagonal terms are flat on the physical side and do not produce a
competing algebraic branch. The only algebraic contribution occurs when
\(i=j\): then \(d_{ii}=0\), the oscillatory exponential cancels, and
\(E_{ii}^{\rm sing}\sim K_{ii}\int^\infty v^{2a_i}e^{-\epsilon v^2/2}\,dv
\sim K_{ii}\epsilon^{-(2a_i+1)/2}\). If all leading breakpoints have the
same power \(a_i=a\), and the leading diagonal coefficient is nonzero, the
algebraic singularity reduces to
\(
K\left(
1-\frac{\Delta^2}{(\Delta_0+2/k_0)^2}
\right)^{-a-\frac12}.
\)

\subsection{Smooth activation functions with $\beta>2$}
For \(\beta>2\), we have \(C_n\to\infty\) in
Eq.~\eqref{eq:saddle_point}. Since \(0<x_n<1\), this equation implies
\(
(1-x_n)^{\beta/2}=\frac{x_n}{C_n}<\frac{1}{C_n},
\)
and hence \(x_n\to1\).
Looking at the terms in Eq \ref{eq:log_fact} we notice that \(\ln(1-x_n)\to-\infty\), the second term, \(\left(1-\frac{2}{\beta}\right)x_n\), stays bounded and, $\frac{1}{2n}(\log(1-x_n)-\log\bigl(2+(\beta-2)x_n\bigr))=O(n^{-1}\log(n))\to 0$. Therefore, substituting the result into Eq.~\eqref{eq:log_fact},
\begin{equation}
\lim_{n\to\infty}
\frac{1}{2n}
\log\left(\frac{|A_{2n}|^2}{(2n)!}\right)
=
-\infty .
\end{equation}
Using Eq.~\eqref{eq:CH}, we obtain \(1/R=\Delta_0^{-1}e^{-\infty}=0\), and hence \(R=\infty\).

\subsection{Non-Smooth Activation Functions}

If the activation function $\phi(x)$ is not infinitely differentiable, the trajectory exhibits finite regularity. Specifically, if the effective potential $V$ is only $C^{q}$ at $\Delta = \Delta_0$, then the force $-\partial_\Delta V$ is $C^{q-1}$. Standard ODE theory guarantees that the solution to $\ddot{\Delta} = -\partial_\Delta V(\Delta)$ is at least $C^{q+1}$ in time, but generally not smoother. 

In this subsection, we demonstrate that if the $q$-th derivative of the activation function, $\phi^{(q)}(x)$, possesses jump discontinuities, then the potential satisfies $V \in C^{q+1}$ but $V \notin C^{q+2}$. This directly causes the temporal trajectory to be $\Delta \in C^{2q+2}$ but $\Delta \notin C^{2q+3}$. As a result, the power spectral density decays asymptotically as a power law,
\begin{equation}
    \tilde \Delta(\omega) \sim \mathcal{O}\bigl(\omega^{-(2q+4)}\bigr).
\end{equation}
This decay is sufficient for the high-frequency part of the Paley--Wiener
integral \eqref{eq:payley_weiner} to converge, and therefore implies
unpredictability in the Paley--Wiener sense.

The physical origin of this power law can be traced to the local singularity
produced by the jump discontinuities of $\phi^{(q)}$. For convenience, we
work with the normalized correlation
\(
    \rho = \Delta/\Delta_0,
\)
so that $\rho \to 1^-$ as $\tau \to 0$.

First consider the behavior of the effective force near $\rho=1$. Suppose
that $\phi^{(q)}(x)$ has jump discontinuities of size $J_i$ at distinct
points $x_i$. In the distributional sense,
\begin{equation}
    D^{q+1}\phi
    =
    \mu_{\rm reg}
    +
    \sum_i J_i \delta_{x_i}.
\end{equation}
Applying Price's theorem ( Gaussian integration by parts),
$q+1$ times gives
\begin{equation}
    \frac{d^{q+1}}{d\rho^{q+1}}
    \mathbb{E}[\phi(X)\phi(Y)]
    =
    \iint p_\rho(x,y)\,
    D^{q+1}\phi(x)\,
    D^{q+1}\phi(y)\, dx\,dy,
\end{equation}
where $X,Y\sim\mathcal{N}(0,\Delta_0)$ where
\begin{equation}
    p_\rho(x,y)
    =
    \frac{1}{2\pi \Delta_0\sqrt{1-\rho^2}}
    \exp\left[
    -\frac{x^2-2\rho xy+y^2}
    {2\Delta_0(1-\rho^2)}
    \right].
\end{equation}

The singular part of $V^{(q+2)}(\rho)$ is therefore controlled by
\(
    \sum_{i,j} J_iJ_j p_\rho(x_i,x_j).
\)
As $\rho\to 1^-$, the off-diagonal terms with $i\neq j$ give exponentially small terms
\(
    \exp\left[
    -\frac{(x_i-x_j)^2}{4\Delta_0(1-\rho)}
    \right].
\)
The diagonal terms $i=j$, give algebraic singularities.
Namely,
\begin{equation}
    p_\rho(x_i,x_i)
    \sim
    C_i(1-\rho)^{-1/2},
    \qquad
    C_i =
    \frac{e^{-x_i^2/(2\Delta_0)}}{2\pi \Delta_0\sqrt{2}} .
\end{equation}
Hence
\(
    V^{(q+2)}(\rho)
    =
    V^{(q+2)}_{\rm reg}(\rho)
    +
    A(1-\rho)^{-1/2}
    +
    \text{exponentially small terms}.
\)
After integrating this relation $q+1$ times with respect to $\rho$, all
analytic integration constants are absorbed into the regular part, leaving
\begin{equation}
    \partial_\rho V(\rho)
    =
    V'_{\rm reg}(\rho)
    +
    B(1-\rho)^{q+\frac12}
    +
    o\!\left((1-\rho)^{q+\frac12}\right).
\end{equation}

The next ingredient is the behavior of the temporal trajectory near
$\tau=0$. Its curvature at the origin is
given by the spectral moment
\begin{equation}
\label{eq:second_derivative}
    \ddot{\Delta}(0)
    =
    -\int_{\mathbb{R}} \omega^2\, d\tilde \Delta(\omega).
\end{equation}
Since the spectral measure is nonnegative, this moment can vanish only if the
spectral weight is supported at $\omega=0$. Thus, for a non-static
trajectory,
\(
    \ddot{\Delta}(0)<0.
\)
Consequently,
\begin{equation}
    1-\rho(\tau)
    =
    -\frac{\ddot{\Delta}(0)}{2\Delta_0}\tau^2
    +
    o(\tau^2).
\end{equation}

Substituting this local behavior into the singular force gives
\(
    (1-\rho(\tau))^{q+\frac12}
    \sim
    |\tau|^{2q+1}.
\)
Since the covariance obeys the equation of motion Eq \ref{eq:newtonian}
the same non-analytic contribution appears in $\ddot{\Delta}(\tau)$. Two
time integrations then give the leading singular part of the trajectory,
\begin{equation}
    \Delta(\tau)
    =
    \Delta_{\rm reg}(\tau)
    +
    c|\tau|^{2q+3}
    +
    o(|\tau|^{2q+3}).
\end{equation}

Finally, a local cusp of the form $|\tau|^{2q+3}$ produces a Fourier tail
\(
    \tilde \Delta(\omega)
    \sim
    \mathcal{O}\bigl(|\omega|^{-(2q+4)}\bigr).
\)

\vspace{1em}
Note that a non-smooth activation function can be made exactly predictable by Gaussian smoothing. Indeed, convolution with the Gaussian kernel $(e^{-|s-t|^2/(2\sigma^2)})$ corresponds, on the Fourier side, to multiplication by \(e^{-\sigma^2\omega^2/2}\). Hence the polynomial high-frequency tail generated by the non-smooth singularity is cut off by a Gaussian factor. This makes Paley--Wiener integral to diverge to \(-\infty\), so the smoothed process becomes exactly predictable.

\section{Complex-Time Analyticity of the Autocorrelation Function $\Delta(\tau)$}

In this section, we analyze the analyticity of the autocorrelation function $\Delta(\tau)$ given the three classes of effective potentials, $V(\Delta)$, described in the previous section. According to the standard theory of ordinary differential equations~\cite{teschl2012ordinary}, if the force $-\partial_{\Delta} V(\Delta)$ is holomorphic in a simply connected region $\Omega$, then $\Delta(\tau)$ is also holomorphic as long as the trajectory remains within $\Omega$, $\Delta(\tau)\in \Omega$. For the potentials considered here, $\Omega$ is the complex plane $\mathbb{C}$ minus branch cuts extending from the singularities of $V(\Delta)$ to infinity. Because the singularities of $\Delta(\tau)$ cannot depend on the arbitrary choice of the deformable branch cuts, the non-analyticities of $\Delta(\tau)$ must be exclusively attained 
at the singular points of the potential $V(\Delta)$  or when \(|\Delta|\to\infty\).

To systematically locate these singularities and determine the analytic structure of $\Delta(\tau)$, we organize the remainder of this section based on the 
asymptotic parameter $\beta$ that controls the Fourier decay rate of the nonlinearities. 
In the first subsection, we employ the complex time-of-flight integral to study the regime $\beta \geq 2$. We demonstrate that the singularities of $\Delta(\tau)$ arise from the trajectory reaching either infinity ($\beta > 2$) or a finite branch point ($\beta = 2$) in finite complex time, and we identify which of these governs the exponential decay rate. In the second subsection, we turn to the regime $0 < \beta < 2$. By examining the essential singularity at the initial condition $\Delta_0$, we show via proof by contradiction that the trajectory is strictly nonanalytic at the origin $\tau=0$, necessitating a trans-series representation of the solution.

\subsection{Analyticity of $\Delta(\tau)$ via the time-of-flight integral ($\beta\geq2$)}

Using energy conservation for the autocorrelation, $\frac{1}{2}\dot{\Delta}(\tau)^2 + V(\Delta(\tau)) = V(\Delta_0)$, we define the complex time-of-flight integral as
\begin{equation}
    \tau(\Delta) = \int_{\Delta_0}^{\Delta}\frac{d \Delta'}{\sqrt{2(V(\Delta_0)-V(\Delta'))}}.
\end{equation}
Once a branch of the square root is chosen, the holomorphic inverse function theorem guarantees that $\Delta(\tau)$ is analytic wherever $V(\Delta)$ is analytic, provided $\tau'(\Delta) \neq 0$. The only points requiring separate analysis are classical turning points ($V(\Delta) = V(\Delta_0)$) and singularities of $V(\Delta)$. At a turning point, assuming $V'(\Delta_0) \neq 0$, the time-of-flight scales as $\tau(\Delta) \sim (\Delta-\Delta_0)^{1/2}$. This inverts to a smooth quadratic minimum $\Delta(\tau) - \Delta_0 \sim \tau^2$, preserving analyticity. Consequently, we arrive at a central conclusion: \emph{singularities in $\Delta(\tau)$ arise exclusively when a singularity of the potential $V(\Delta)$ is reached in a finite complex time.}
We show below that this finite-time condition is satisfied for the potentials considered here.

\paragraph{Entire potentials ($\beta>2$):}
For $\beta >2$, $V(\Delta)$ has no singularities at finite $\Delta$; however, its singularity at infinity causes a finite-time singularity in $\Delta(\tau)$. As shown in the previous section, for $\beta>2$, $V(\Delta)$ is an entire function and can be expanded as
\begin{equation}
    V(\Delta) = \sum_{n\ge0} a_n \Delta^n \ge \frac{a_N}{N+1}\Delta^{N+1},
\end{equation}
with $a_n>0$ and an integer $N > 2$. Consequently, for sufficiently large \(\Delta\), we have \(|V(\Delta_0)-V(\Delta)|\ge C'\Delta^{N+1}\), which bounds the time-of-flight integrand:
\begin{equation}
    \left| \frac{1}{\sqrt{2(V(\Delta_0)-V(\Delta))}} \right| \le \tilde{C}\Delta^{-(N+1)/2}.
\end{equation}
Because \(N\ge 2\), the integral converges at infinity: \(\int^\infty \Delta^{-(N+1)/2}\,d\Delta < \infty\). Thus, the trajectory reaches infinity at a finite complex time, manifesting as a singularity in $\Delta(\tau)$.

\paragraph{Finite branch points ($\beta = 2$):}
For $\beta=2$, the potential \(V(\Delta)\) has a finite branch-point singularity at $S > \Delta_0$. Defining \(u = S-\Delta\), we may write the potential near the singularity as
\begin{equation}
    V(S-u) = V_{\rm reg}(S-u) + B u^\alpha, \qquad B\neq0,
\end{equation}
where \(V_{\rm reg}\) is analytic at \(u=0\). Let \(V_* \equiv V_{\rm reg}(S)\). To determine if the singularity is reached, we evaluate whether the local time-of-flight integral
\begin{equation}
    \int_0^\epsilon \frac{du}{\sqrt{2(V(\Delta_0)-V(S-u))}}
\end{equation}
is finite. 

If \(\alpha>0\) and \(V(\Delta_0)\neq V_*\), the denominator remains finite and non-zero near \(u=0\), making the integrand bounded. Hence, the local time-of-flight is trivially finite. If \(\alpha<0\), the singular term dominates, such that \(V(\Delta_0)-V(S-u) \sim -B u^\alpha\). The integrand then scales as
\begin{equation}
    \frac{1}{\sqrt{2(V(\Delta_0)-V(S-u))}} \sim \frac{1}{\sqrt{-2B}}u^{-\alpha/2}.
\end{equation}
Because \(\alpha<0\), the exponent \(-\alpha/2\) is positive, ensuring that \(\int_0^\epsilon u^{-\alpha/2}\,du < \infty\). In both cases, the local time-of-flight is finite, meaning the obstruction is reached at a finite complex time.

To accurately characterize the exponential decay rate of the autocorrelation, we must identify the singularity of $\Delta(\tau)$ nearest to the real axis. For $\beta=2$, $\Delta(\tau)$ possesses two potential sources of singularity: the finite singularity of $V(\Delta)$ at $S$, and the asymptotic behavior $\lim_{|\Delta| \to \infty}V(\Delta)$. To find the correct decay rate, we must determine which of these contributes to the nearest singularity in complex time. 

As stated in the main text, for many practical transfer functions (e.g., the error function and the smoothed ReLU), the potential is bounded along the relevant large-\(\Delta\) branches by $|V(\Delta)|\leq C|\Delta|^2\log(|\Delta|)$. Both of these examples are governed by the arcsine singularity of the Gaussian sign covariance,
\begin{equation}
    \mathbb{E}[\operatorname{sgn}X\,\operatorname{sgn}Y] = \frac{2}{\pi}\arcsin\left(\frac{\Delta}{q}\right).
\end{equation}
The erf nonlinearity is a Gaussian-smoothed sign, while the smoothed ReLU is an integrated smoothed step. By Price's theorem, their relevant covariances are obtained by integrating the same arcsine kernel, yielding \(C_{\rm erf}(\Delta)=\mathcal{O}(\log\Delta)\) and \(C_{\rm s-ReLU}(\Delta)=\mathcal{O}(\Delta\log\Delta)\). This results in an effective potential bounded by \(V_{\rm s-ReLU}(\Delta)=\mathcal{O}(\Delta^2\log\Delta)\). 

Under this bound, the time required to reach infinity diverges:
\begin{equation}
     \int^\infty \frac{d\Delta}{\sqrt{2(V(\Delta_0)-V(\Delta))}} \sim \int^\infty \frac{d\Delta}{\sqrt{\Delta^2 \log\Delta}} = 2\sqrt{\log \Delta}\bigg|^\infty = \infty.
\end{equation}
Because infinity is never reached in finite time, it does not generate a singularity in $\Delta(\tau)$. Therefore, the exponential tail is strictly determined by the time of flight to the finite singularity of the potential at $S$.
 
\subsection{Analyticity of $\Delta(\tau)$ for $0 < \beta <2$} 

In this subsection, we demonstrate that the singular structure of the potential $V(\Delta)$ at the initial condition $\Delta_0$ implies that the trajectory $\Delta(\tau)$ is strictly nonanalytic at $\tau=0$. We work with the dimensionless local coordinate at the
branch point,
\begin{equation}
x=\frac{\Delta-\Delta_0}{\Delta_0} .
\label{eq:x_def}
\end{equation}
Since \(\Delta_0\) is the real turning point, the physical trajectory approaches the branch point from \(x<0\). The opposite side, \(x>0\), is reached only by analytic continuation. From the previous subsection, crossing to this continued side produces two lateral branches of the potential whose difference is exponentially small. To keep track only of its leading exponential
scale, we write
\(
    F \doteq e^{-\Phi}
\)
if, in the relevant singular limit,
\(
    \log |F| \sim -\Phi 
\) to the leading order on x.
Thus algebraic prefactors are suppressed. With this convention,
\begin{equation}
    V_+(x)-V_-(x)
    \doteq
    \exp[-Kx^{-q}],
    \qquad
    x\to0^+,
    \label{eq:DiscV_general}
\end{equation}
where  \(K>0\), \(q=\beta/(2-\beta)\) and \(V_\pm\) are the two lateral continuations of the potential across the
branch cut starting from \(x=0\). We now track how this obstruction is seen by the trajectory itself. To do so,
we substitute the continued trajectory into the integral representation for
the force.
We separate the local solution asymptotically as
\begin{equation}
x(\tau)
=
x_{\rm reg}(\tau)
+
x_{\rm sing}(\tau),
\label{eq:Delta_split}
\end{equation}
where \(x_{\rm reg}\) is the analytic Taylor part, while
\(x_{\rm sing}\) denotes the contribution associated
with the branch-cut obstruction.
 For the analytic part of the trajectory, the local coordinate has the expansion
\[
    x_{\rm reg}(\tau)
    =
    -\frac{C}{\Delta_0}\tau^2+O(\tau^4),
    \qquad C>0 .
\]
Here \(C>0\) follows from the covariance property of \(\Delta(\tau)\) (Eq \ref{eq:second_derivative}), as
discussed above.
For real \(\tau\),this equation gives
\(x_{\rm reg}(\tau)<0\), so the trajectory remains on the physical side,
where the potential is real and smooth. However, analyticity at
\(\tau=0\) would require a holomorphic continuation in complex \(\tau\).
Along the imaginary direction, \(\tau=is\), one obtains
\( x_{\rm sing}(is)
\approx
\frac{C}{\Delta_0}s^2+O(s^4)
\)
Thus the imaginary \(\tau\)-direction maps the turning point to the
continued side of the \(x\)-plane, where the two analytic continuations of
the potential differ by the discontinuity \eqref{eq:DiscV_general}.

Substituting the continued trajectory into the integral representation for
the force Eq \ref{eq:assymptotics_beta} gives
\begin{equation}
\partial_\Delta V(\Delta(is))
\sim
\int_0^\infty
\frac{
\exp\left[-B t^{-q}\right]
}
{1-\left(1+x_{\rm reg}(is)+x_{\rm sing}(is)\right)^2 e^{-t}}
\,dt .
\label{eq:Eprime_pullback}
\end{equation}
On the continued side, the denominator has a pole at
\(
t_0(s)
=
\log\left(1+x_{\rm reg}(is)+x_{\rm sing}(is)\right)^2 .
\)
We first work under the provisional estimate
\(
    |x_{\rm sing}(is)|
    \leq
    M e^{-\kappa_0 s^{-2q}}
\),where \(\kappa_0\) will be chosen below.
This estimate will be verified a posteriori. Hence the pole
location is, to the leading accuracy, determined by the regular
trajectory:
\[
t_0(s)
=
\log\left(1+x_{\rm reg}(is)+x_{\rm sing}(is)\right)^2
\sim
\frac{2C}{\Delta_0}s^2 .
\]
The two lateral continuations differ by the residue at this pole. Evaluating the numerator at \(t=t_0(s)\) gives
discontinuity of the size \(
\exp[-B t_0(s)^{-q}]
\sim
\exp[-\kappa s^{-2q}],
\)
with
\(
\kappa
=
B\left(\frac{\Delta_0}{2C}\right)^q>0 .
\)
Depending on which side of the branch cut is chosen, the continued
trajectory satisfies one of the two equations
\(
\ddot{\Delta}_\pm
=
-\partial_\Delta V_\pm(\Delta_\pm).
\)
We denote the corresponding discontinuity of the trajectory by
\begin{equation}
\operatorname{Disc}_{\tau}\Delta
:=
\Delta_+(\tau)-\Delta_-(\tau).
\end{equation}
Since the two branches have the same regular part, the leading
nonperturbative discontinuity satisfies
\begin{equation}
\partial_\tau^2\operatorname{Disc}_{\tau}\Delta
\doteq
\exp\left[-\frac{\kappa}{\tau^m}\right],
\qquad
m=2q=\frac{2\beta}{2-\beta}.
\label{eq:DiscDelta_force}
\end{equation}
Integrating \eqref{eq:DiscDelta_force} does not change the exponential
scale. To see this, one may include an arbitrary algebraic prefactor and use the
large-argument asymptotics of the upper incomplete gamma function:
\[
\int_0^\tau s^{-p}e^{-\kappa/s^m}\,ds
=
\frac{1}{m}\kappa^{\frac{1-p}{m}}
\Gamma\left(\frac{p-1}{m},\frac{\kappa}{\tau^m}\right).
\]
Since \(\Gamma(a,x)\sim x^{a-1}e^{-x}\) for \(x\to\infty\), the integral is
still of scale \( e^{-\kappa/\tau^m}\).After two integrations,
\(
\operatorname{Disc}_{\tau}\Delta
\doteq
\exp\left[-\frac{\kappa}{\tau^m}\right].
\label{eq:DiscDelta_final}
\)
Consequently, along \(\tau=is\),
\[
|x_{\rm sing}(is)|
\leq
M' \exp[-\kappa s^{-m}],
\qquad m=2q .
\]
We now choose \(\kappa_0\) so that
\(
    0<\kappa_0<\kappa .
\)
With this choice, the estimate obtained above improves the provisional
estimate. Thus the smallness assumption on \(x_{\rm sing}\) is
self-consistent. At a more rigorous
level, this can be phrased as a bootstrap/continuation argument: the set of
\(s\)-values for which the provisional bound holds is nonempty, open by
continuity, and closed because the improved estimate gives a fixed margin, therefore this set coincides with the
whole interval.

In summary, the two continued branches of the trajectory may be written as
\begin{equation}
\Delta_\pm(\tau)
=
\Delta_{\rm reg}(\tau)
\pm
\text{terms of exponential scale }
\exp\left[-\frac{\kappa}{\tau^m}\right],
\qquad
m=\frac{2\beta}{2-\beta}.
\end{equation}
Thus the full continued trajectory has zero radius of analyticity at
\(\tau=0\) and its two lateral continuations differ by an exponentially
small branch jump of scale \(\exp[-\kappa/\tau^m]\).

We now show that this local asymptotic form produces a stretched-exponential contribution to \(\tilde \Delta(\omega)\) at large \(\omega\), with stretched exponent \(m/(m+1)\).
For \(\omega>0\), deform the real Fourier contour into the lower half-plane,
where the factor \(e^{-i\omega t}\) decays. Assume that the leading obstruction is a branch contour \(\Gamma\) attached to
\(t=0\). The corresponding leading asymptotic contribution is then
\[
\tilde \Delta(\omega)
\sim
I_\Gamma(\omega),
\qquad
I_\Gamma(\omega)
=
\int_\Gamma \tilde \Delta_\Gamma(t)e^{-i\omega t}\,dt .
\]
Using the local form
 \(\tilde  \Delta\), we obtain
\(
I_\Gamma(\omega)
\sim
S\int_\Gamma
t^\beta
\exp\left(
-\frac{\kappa}{t^m}-i\omega t
\right)
\left(1+\sum_{k=1}^{\infty}a_k t^k\right)\,dt .
\)

The relevant scale is found by balancing the two terms in the exponent:
\(
\frac{\kappa}{t^m}\sim \omega t .
\)
We therefore set
\(
t=\omega^{-1/(m+1)}z .
\)
Hence
\begin{equation}
I_\Gamma(\omega)
\sim
S\,
\omega^{-(\beta+1)/(m+1)}
\int_{\omega^{1/(m+1)}\Gamma}
z^\beta
\exp\left[
-\omega^{m/(m+1)}
\left(
\frac{\kappa}{z^m}+iz
\right)
\right]dz .
\end{equation}
Define
\(
\lambda=\omega^{m/(m+1)},
\)\(
\phi(z)=\frac{\kappa}{z^m}+iz .
\)
The saddle points satisfy
\(
-\frac{m\kappa}{z^{m+1}}+i=0 .
\) so
\(
\phi''(z)=\frac{m(m+1)\kappa}{z^{m+2}},
\)

The saddle point approximation:
\(
\int z^\beta e^{-\lambda\phi(z)}\,dz
\sim
z_*^\beta
\left(
\frac{2\pi}{\lambda\phi''(z_*)}
\right)^{1/2}
e^{-\lambda\phi(z_*)}
\)
gives
\begin{equation}
I_\Gamma(\omega)
\sim
C\,
\omega^{-\frac{2\beta+m+2}{2(m+1)}}
\exp\left(
-\Lambda\omega^{m/(m+1)}
\right),
\end{equation}
where
\(
\Lambda=\phi(z_*) .
\)
Since \(m/(m+1)<1\), the stretched-exponential tail
\(\exp[-\Lambda \omega^{m/(m+1)}]\) is not fast enough to make the
Paley--Wiener integral diverge to \(-\infty\).
Thus the integral remains finite, and the process is not perfectly
predictable.

\section{Equivalence of the PSD and the LDOS on Krylov chain}

We briefly recall the correspondence between orthogonal polynomials, Jacobi matrices, and scalar spectral measures. 
The purpose is to make explicit how the prescribed density \(\tilde\Delta(\omega)\) is realized as the local density of states (LDOS) of a Lanczos chain at \(e_0\), and how this construction gives a stationary Gaussian process with the prescribed spectral density. The point is standard, but since we proceed in the inverse direction---from the prescribed spectral density to the Lanczos chain---we spell out the density condition needed for the scalar spectral measure of the chain to coincide with \(\tilde\Delta(\omega)d\omega\). 

For simplicity we assume \(\int_{\mathbb R}\tilde\Delta(\omega)d\omega=1\);
otherwise one applies the construction to the normalized measure and rescales the
resulting Gaussian process by \(\sqrt{\Delta(0)}\).
Recall that the recurrence coefficients of the orthogonal polynomials associated
with the measure \(\tilde\Delta(\omega)\,d\omega\) define a Jacobi matrix, or
equivalently a Lanczos chain, \(J\).
For a finite chain, if \(J\psi_\alpha=E_\alpha\psi_\alpha\), the LDOS at \(e_0\) is
\(
    \rho_J(\omega)\,d\omega
    =
    \sum_\alpha |\langle e_0,\psi_\alpha\rangle|^2
    \delta(\omega-E_\alpha)\,d\omega .
\)
Equivalently, since
\(
    f(J)=\sum_\alpha f(E_\alpha)
    |\psi_\alpha\rangle\langle\psi_\alpha| ,
\)
testing the LDOS against a bounded function \(f\) gives
\begin{equation}
    \int_{\mathbb R} f(\omega)\rho_J(\omega)\,d\omega
    =
    \sum_\alpha |\langle e_0,\psi_\alpha\rangle|^2 f(E_\alpha)
    =
    \langle e_0,f(J)e_0\rangle .
\end{equation}
In the infinite-chain limit, this identity defines the LDOS \(\rho_J(\omega)\) at \(e_0\).

By Theorem~2.23 in Teschl \cite{teschl2000jacobi}, if the closed linear span of the
orthogonal polynomials is the whole Hilbert space
\(L^2(\mathbb R,\tilde\Delta(\omega)d\omega)\), the space of square-integrable
functions with respect to the measure \(\tilde\Delta(\omega)d\omega\), then the Jacobi operator
constructed from their recurrence coefficients has
\(\tilde\Delta(\omega)d\omega\) as its LDOS measure at \(e_0\).
Taking \(f_t(J)=e^{-iJt}\), this gives
\begin{equation}
\begin{aligned}
\mathcal F_\tau\!\left[
\langle e_0,e^{-iJ\tau}e_0\rangle
\right](\omega)
&=
\int_{\mathbb R}\tilde\Delta(\lambda)
\left[
\int_{\mathbb R}e^{i(\omega-\lambda)\tau}\,d\tau
\right]d\lambda 
&=
2\pi\tilde\Delta(\omega).
\end{aligned}
\end{equation}
Thus the \(e_0\)-projected return amplitude is the time-domain correlator whose
Fourier transform gives the LDOS. The factor \(2\pi\) only reflects the Fourier
convention.
\paragraph{Density of the polynomial subspace}
It remains to justify that the polynomial subspace spans the whole Hilbert space.
In our setting this is automatic. Let
\(m_n=\int_{\mathbb R}\omega^n\tilde\Delta(\omega)\,d\omega\). Since
\(\Delta(\tau)=\int_{\mathbb R}e^{-i\omega\tau}\tilde\Delta(\omega)\,d\omega\), we
have \(\Delta^{(n)}(0)=(-i)^n m_n\). A sufficient condition for this  is the Carleman's condition\cite{SIMON199882},
\(\sum_{n=1}^{\infty}m_{2n}^{-1/(2n)}=\infty\), where
\(m_{2n}=\int_{\mathbb R}\omega^{2n}\tilde\Delta(\omega)\,d\omega\). The nonzero radius of convergence
\(R>0\) of \(\Delta(\tau)\) at the origin therefore implies
\(R^{-1}=\limsup_{n\to\infty}|m_n/n!|^{1/n}<\infty\). Hence, for any
\(A>R^{-1}\), \(m_{2n}\le A^{2n}(2n)!\) for all sufficiently large \(n\).
Taking inverse \(2n\)-th roots gives
\(m_{2n}^{-1/(2n)}\ge A^{-1}((2n)!)^{-1/(2n)}\sim e/(2An)\), where the last step
uses Stirling's formula. Therefore
\(\sum_{n=1}^{\infty}m_{2n}^{-1/(2n)}=\infty\), so Carleman's condition holds.
Consequently, the closed linear span of the polynomials is
\(L^2(\mathbb R,\tilde\Delta(\omega)d\omega)\), as required.
\paragraph{Gaussian process}
We now use this Jacobi chain as a linear realization of the Gaussian process.
Choose white Gaussian initial Krylov coordinates
\(
|s(0)\rangle=\sum_{k\ge0}s_k(0)|e_k\rangle,
\)
\(
\mathbb E[s_k(0)\overline{s_\ell(0)}]=\delta_{k\ell}.
\)
The Krylov state evolves as\\
\(
|s(t)\rangle=e^{-iJt}|s(0)\rangle .
\)
The observed process is the boundary coordinate
\(
X_{\rm K}(t):=\operatorname{Re}\langle e_0|s(t)\rangle
=
\operatorname{Re}\langle e_0|e^{-iJt}|s(0)\rangle .
\)

Its autocorrelation is
\begin{equation}
\begin{aligned}
\mathbb E\!\left[X_{\rm K}(t)\overline{X_{\rm K}(0)}\right]
=
\mathbb E\!\left[
    \langle e_0|e^{-iJt}|s(0)\rangle
    \overline{\langle e_0|s(0)\rangle}
\right] =
\langle e_0|e^{-iJt}\,
\mathbb E[|s(0)\rangle\langle s(0)|]
|e_0\rangle =
\langle e_0|e^{-iJt}|e_0\rangle .
\end{aligned}
\end{equation}
Therefore the Fourier transform of the autocorrelation is the LDOS:
\begin{equation}
S_{X_{\rm K}}(\omega)
=
\mathcal F_t\!\left[
    \mathbb E[X_{\rm K}(t)X_{\rm K}(0)]
\right](\omega)
=
2\pi\rho_0(\omega)
=
2\pi\tilde\Delta(\omega),
\end{equation}
with the Fourier convention above. Thus, the time evolution of the \(e_0\)-coordinate, initialized with i.i.d. standard Gaussian data, has the same PSD as the original Gaussian process. Since a centered stationary Gaussian process is determined by its PSD, the Jacobi chain provides a linear realization of that process.

\section{Dependence of the complex time singularity on the smoothening Gaussian}
As discussed above, for nonlinearities whose Fourier transform has a
Gaussian tail, the singularity of the potential emerges at
\begin{equation}
    R=\Delta_0+\frac{2}{k_0}.
\end{equation}
For Gaussian smoothing, convolution in real space corresponds to multiplication
by \(e^{-\sigma^2\omega^2/2}\) in Fourier space. Since nonsmooth nonlinearities
produce only polynomial large-\(\omega\) Fourier tails, the exponential
contribution comes solely from this Gaussian factor. Thus \(2/k_0=\sigma^2\),
and therefore
\(
    R=\Delta_0+\sigma^2 .
\) So large assymptotic of $t_s$ is given by
    \begin{equation}
	t_s
	=\int_{\Delta_0}^{\Delta_0+\sigma^2}
	\frac{d\Delta}{\sqrt{2(V(\Delta_0)-V(\Delta))}}\sim \sigma\sqrt{\frac{2}{-V'(\Delta_0)}}
	\end{equation}
so $\alpha=\frac{\pi}{2 t_s}\sim\frac{\pi}{2\sigma}\sqrt{\frac{V'(\Delta_0)}{2}}$, and hence it diverges as \(\sigma^{-1}\) in the
\(\sigma\to0\) limit.

\section{Complex Time Singularity and Asymptotic Scaling}

In this section, we analyze the complex time singularity $t_S$ as a function of the coupling $g$. 
For a transfer function with a Gaussian Fourier tail or faster ($\beta \geq 2$), $t_S$ is governed by the integral
\begin{equation}
    t_S = -i \int_{\Delta_0}^{S} \frac{d\Delta}{\sqrt{2[V(\Delta_0)-V(\Delta)]}} \equiv \int_{\Delta_0}^{S} \frac{d\Delta}{\sqrt{V(\Delta)}},
\end{equation}
where $S \geq \Delta_0$ is the location of the singularity of the effective potential $V(\Delta)$. 
The second equality is obtained by adopting the convention that $V(\Delta_0) = 0$ and absorbing the factor of $2$ into the definition of $V(\Delta)$. 
Note that the domain of integration extends beyond the physical regime ($|\Delta| \leq \Delta_0$), and $V(\Delta) \geq V(\Delta_0) = 0$ throughout this region.   

Energy conservation, combined with the asymptotic boundary condition $\lim_{\tau \to \infty}\dot{\Delta}(\tau) = 0$, implies that $V(\Delta_\infty) = 0$. Consequently, the effective potential $V(\Delta)$ possesses roots at both $\Delta = \Delta_0$ and $\Delta = \Delta_\infty$. Furthermore, the value of $\Delta_\infty$ dictates the underlying symmetry of $V(\Delta)$: if $\Delta_\infty = 0$, $V(\Delta)$ is strictly even in $\Delta$, whereas for $\Delta_\infty \neq 0$, this symmetry is broken.

\subsection{Scaling near the critical point}

Near criticality, we rely on standard normal-form analysis: a smooth family of functions with nearby roots is well-approximated in the vicinity of these roots by the lowest-degree root-preserving polynomial compatible with the symmetries of the system~\cite{murdock2003normal}. 
For a symmetric potential ($\Delta_\infty = 0$), the normal form is $V_4(\Delta) = K(0) \Delta^2 (\Delta^2 -\Delta_0^2)$, while for the asymmetric case ($\Delta_\infty \neq 0$), it is $V_3(\Delta) = K(0) (\Delta - \Delta_\infty)^2 (\Delta -\Delta_0)$. 

Because $V(\Delta)$ is analytic in the integration region, we can extract the leading divergence by splitting the integral at an intermediate matching scale $\Delta_c$, chosen such that $\Delta_0 \ll \Delta_c \ll S$. This introduces two competing sources of error: an inner error $\epsilon_{\mathrm{in}}$ due to the neglect of higher-order terms in the normal form, and an outer error $\epsilon_{\mathrm{out}}$ due to the truncation of the integration domain. We decompose the integral as
\begin{equation}
    t_S = \int_{\Delta_0}^{\Delta_c} \frac{d\Delta}{\sqrt{V_p(\Delta)}} + \epsilon_{\mathrm{in}}(\Delta_c) + \epsilon_{\mathrm{out}}(\Delta_c),
\end{equation}
where $p \in \{3,4\}$ is dictated by symmetry. The respective errors are defined as
\begin{align}
    \epsilon_{\mathrm{in}}(\Delta_c) &= \int_{\Delta_0}^{\Delta_c} \left( \frac{1}{\sqrt{V(\Delta)}} - \frac{1}{\sqrt{V_p(\Delta)}} \right) d\Delta, \\
    \epsilon_{\mathrm{out}}(\Delta_c) &= \int_{\Delta_c}^S \frac{d\Delta}{\sqrt{V(\Delta)}}.
\end{align}

We first analyze the inner region, $1 \le \Delta/\Delta_0 \le \Delta_c/\Delta_0$. In the regime $\Delta_0 \ll \Delta \ll S$, the prefactor $K(\Delta)$ can be expanded as $K(\Delta) = K(0) + \mathcal{O}(\Delta/S)$. This follows from the fact that $K'(\Delta) \sim f'(0)/S$, where $f(x) = K(\Delta/S)$ is Lipschitz continuous near the origin. Consequently, the ratio of the true potential to the normal form is
\begin{equation}
    \frac{V(\Delta)}{V_p(\Delta)} = 1 + \mathcal{O}\left(\frac{\Delta}{S}\right).    
\end{equation}
This guarantees that non-polynomial corrections are uniformly suppressed for $\Delta_0 \ll \Delta \ll S$, and $|K(0) - K(\Delta_c)|$ remains small. 

Substituting the normal forms into the leading-order integral yields the dominant divergent behavior of $t_S$. For the symmetric case ($p=4$), we obtain
\begin{align}
    \int_{\Delta_0}^{\Delta_c} \frac{d\Delta}{\sqrt{K(0)}\Delta\sqrt{\Delta^2-\Delta^2_0}} 
    &= \frac{1}{\Delta_0\sqrt{K(0)}} \arccos\left(\frac{\Delta_0}{\Delta_c}\right) \nonumber \\
    &\approx \frac{\pi}{2\Delta_0\sqrt{K(0)}}.
\end{align}
For the asymmetric case ($p=3$), the integral yields
\begin{align}
    \int_{\Delta_0}^{\Delta_c} \frac{d\Delta}{\sqrt{K(0)}\Delta\sqrt{\Delta-\Delta_0}} 
    &= \frac{2}{\sqrt{\Delta_0 K(0)}} \arctan\left(\sqrt{\frac{\Delta_c}{\Delta_0}-1}\right) \nonumber \\
    &\approx \frac{\pi}{\sqrt{\Delta_0 K(0)}}.
\end{align}

The constant $K(0)$ is fixed by the second derivative of the potential at $\Delta = 0$. 
In the limit $\Delta_0 \to 0$, the symmetric case gives $K(0) = 1/(\Delta_0\tau_\infty)^2$, which reproduces the known autocorrelation decay $\Delta(\tau) = \Delta_0 \operatorname{sech}(\tau/\tau_\infty)$~\cite{CrisantiSompolinsky2018} and results in a scaling of $(\pi/2) \tau_{\infty}$. 
Conversely, for the asymmetric case, $K(0) = 1/(\Delta_0\tau_\infty^2)$, which results in the scaling $\pi \tau_{\infty}$. 

To explicitly estimate the inner error $\epsilon_{\mathrm{in}}(\Delta_c)$ incurred by substituting the exact potential with its normal form, we expand the potential to next-to-leading order in the inner region as $V(\Delta) \approx V_p(\Delta)\left(1 + \frac{\Delta}{S}\right)$. The integrand can then be approximated as
\begin{equation}
    \frac{1}{\sqrt{V(\Delta)}} - \frac{1}{\sqrt{V_p(\Delta)}} \approx -\frac{\Delta}{2S\sqrt{V_p(\Delta)}}.
\end{equation}
Using this expansion, the inner error is directly given by the integral of this subleading term:
\begin{equation}
    \epsilon_{\mathrm{in}}(\Delta_c) \approx -\frac{1}{2S} \int_{\Delta_0}^{\Delta_c} \frac{\Delta}{\sqrt{V_p(\Delta)}} \,d\Delta.
\end{equation}

For the symmetric case ($p=4$), substituting $V_4(\Delta)$ yields a logarithmic error estimate:
\begin{align}
    \epsilon_{\mathrm{in}}^{(4)}(\Delta_c) &\approx -\frac{1}{2S\sqrt{K(0)}} \int_{\Delta_0}^{\Delta_c} \frac{d\Delta}{\sqrt{\Delta^2-\Delta_0^2}} \nonumber \\
    &\approx -\frac{1}{2S\sqrt{K(0)}} \ln\left(\frac{2\Delta_c}{\Delta_0}\right),
\end{align}
where the final approximation relies on the scale separation $\Delta_c \gg \Delta_0$. 
To determine the exact asymptotic scaling of the inner errors, we express the constant $K(0)$ in terms of the critical parameter $\Delta_0$, $K(0) = 1/(\tau_\infty \Delta_0)^2$. Substituting this into our logarithmic error estimate yields
\begin{equation}
  \lim_{\Delta_0 \to 0} \frac{\epsilon_{\mathrm{in}}^{(4)}}{\tau_\infty} \approx 
    \lim_{\Delta_0 \to 0} \frac{\Delta_0}{2S} \ln(\Delta_0) = 0.
\end{equation}
Thus, the inner error scales as $o(\tau_\infty)$. 

For the asymmetric case ($p=3$), substituting $V_3(\Delta)$ yields a square-root error estimate:
\begin{align}
    \epsilon_{\mathrm{in}}^{(3)}(\Delta_c) &\approx -\frac{1}{2S\sqrt{K(0)}} \int_{\Delta_0}^{\Delta_c} \frac{d\Delta}{\sqrt{\Delta-\Delta_0}} \nonumber \\
    &\approx -\frac{\sqrt{\Delta_c}}{S\sqrt{K(0)}}.
\end{align}
Putting, $K(0) = 1/(\tau_\infty^2 \Delta_0)$, yields
\begin{equation}
    \lim_{\Delta_0 \to 0} \frac{\epsilon_{\mathrm{in}}^{(3)}}{\tau_\infty} \approx
    -\lim_{\Delta_0 \to 0} \frac{\sqrt{\Delta_c}}{S} \sqrt{\Delta_0} = 0.
\end{equation}
Again, the inner error scales as $o(\tau_\infty)$. In both cases, the inner errors diverge strictly slower than the leading terms as $\Delta_0 \to 0$, rigorously confirming that the lowest-order normal form approximation captures the exact leading critical behavior.

To estimate the outer error $\epsilon_{\mathrm{out}}$ and prove it is strictly subdominant, we turn to the microscopic form of the potential:
\begin{equation}
    V(\Delta) = -\Delta^2 + 2g^2 \int_0^\Delta \langle \phi(x_1) \phi(x_2) \rangle_{s} \,ds.
\end{equation}
Assuming the correlation function can be expanded in a series of polynomials, we write $\langle \phi(x_1) \phi(x_2) \rangle_{s} = \sum_{n} a_n^2 s^n$, with $a_n^2 \geq 0$. We then define the auxiliary function $E(\Delta)$:
\begin{equation}
    E(\Delta) \equiv \frac{2}{\Delta^2} \int_0^\Delta \langle \phi(x_1) \phi(x_2) \rangle_{s} \,ds = \sum_{n\geq 1} \frac{2a_n^2}{n+1} \Delta^{n-1}.  
\end{equation}
The critical condition $V(\Delta_0) = 0$ requires $\Delta_0^2[-1 + g^2 E(\Delta_0)] = 0$, fixing the coupling at $g^2 = 1/E(\Delta_0)$. The potential can therefore be rewritten exactly as
\begin{equation}
    V(\Delta) = \Delta^2 \left[-1 + \frac{E(\Delta)}{E(\Delta_0)} \right].
\end{equation}

Taking the derivative of $E(\Delta)$ yields $E'(\Delta) = \sum_{n\geq1} \frac{2(n-1)}{n+1} a_n^2 \Delta^{n-2}$. Let $m$ denote the index of the leading non-trivial term; this contributes $2\frac{m-1}{m+1} a_m^2 \Delta^{m-2}$. Since all remaining terms in the sum are manifestly non-negative, we can bound $E'(\Delta)$ below by this leading term. Integrating this inequality from $\Delta_0$ to $\Delta$ provides a strict lower bound on $E(\Delta)$:
\begin{equation}
    E(\Delta) - E(\Delta_0) \ge \int_{\Delta_0}^\Delta 2\frac{m-1}{m+1} a_m^2 s^{m-2} \,ds = C(\Delta^{m-1}-\Delta_0^{m-1}),
\end{equation}
where $C = \frac{2}{m+1} a_m^2$. For the symmetric case ($\Delta_\infty = 0$), the leading non-trivial term corresponds to $m=3$, while for the asymmetric case ($\Delta_\infty \neq 0$), it is $m=2$.

Using this bound, we can strictly constrain the outer error:
\begin{align}
    \epsilon_{\mathrm{out}} &= \sqrt{E(\Delta_0)} \int_{\Delta_c}^S \frac{d\Delta}{\Delta \sqrt{E(\Delta) - E(\Delta_0)}} \nonumber \\
    &\leq \sqrt{\frac{E(\Delta_0)}{C}} \int_{\Delta_c}^S \frac{d\Delta}{\Delta\sqrt{\Delta^{m-1} -\Delta_0^{m-1}}} \nonumber \\
    &= \sqrt{\frac{E(\Delta_0)}{C}} \frac{1}{\Delta_0^{(m-1)/2}} \int_{\Delta_c/\Delta_0}^{S/\Delta_0} \frac{dy}{y \sqrt{y^{m-1} -1}}.
\end{align}
For small arguments ($\Delta_0 \ll \Delta_c \ll S$), the integral is approximately:
\begin{equation}
    \epsilon_{\mathrm{out}} \approx C' \frac{1}{\Delta_0^{(m-1)/2}} \left[ \left(\frac{\Delta_0}{\Delta_c}\right)^{(m-1)/2} - \left(\frac{\Delta_0}{S}\right)^{(m-1)/2} \right].
\end{equation}
Crucially, multiplying by $\Delta_0^{(m-1)/2}$ and taking the critical limit yields
\begin{equation}
    \lim_{\Delta_0 \to 0} \left( \epsilon_{\mathrm{out}} \Delta_0^{(m-1)/2} \right) = 0.
\end{equation}
Therefore, $\epsilon_{\mathrm{out}} = o(\Delta_0^{-(m-1)/2})$. 
For the symmetric case ($m = 3$), the error scales as $\mathcal{O}(\Delta_c^{-1})$. For the asymmetric case ($m=2$), the error scales as $o(\Delta_0^{-1/2})$. 
Since the mixing time scales as $\tau_\infty \sim \Delta_0^{-1}$ when $\Delta_{\infty} = 0$, and $\tau_\infty \sim \Delta_0^{-1/2}$ when $\Delta_{\infty} \neq 0$~\cite{EngelkenWolfAbbott2023,CrisantiSompolinsky2018}, the outer error is always strictly bounded by $\epsilon_{\mathrm{out}} = o(\tau_\infty)$. 

Combining these results, the complex time singularity near the critical point exhibits the following exact asymptotic scaling:
\begin{equation}
    t_S = 
    \begin{cases} 
        \frac{\pi}{2} \tau_\infty + o(\tau_\infty), & \text{for } \Delta_\infty = 0 \text{ (symmetric),} \\[1ex]
        \pi \tau_\infty + o(\tau_\infty), & \text{for } \Delta_\infty \neq 0 \text{ (asymmetric).}
    \end{cases}
\end{equation}

While the normal-form approximation is strictly controlled only in the region $\Delta_0 \ll \Delta_c \ll S$, it provides an excellent numerical approximation over a much broader range of couplings $g$ (see Fig.~\ref{fig:ts_asymptotics}), effectively capturing the crossover physics beyond the strict asymptotic regime. The departure of $t_S$ from the leading asymptotic $\tau_\infty$ scaling occurs faster [$\mathcal{O}(\Delta_0)$ vs $\mathcal{O}(\sqrt{\Delta_0})$] for the asymmetric case than for the symmetric case.

\subsection{Scaling in the strong-coupling limit ($g \to \infty$)}

In the strong-coupling limit, the critical overlap scales as $\Delta_0 \sim g^2$. Therefore, for $g \gg 1$, we have $\Delta_0 \gg 1$. As a result, the turning point approaches the strict boundary of the domain, causing the upper limit of the integral to collapse ($S - \Delta_0\ll\Delta_0$). Because the entire integration domain is compressed near $\Delta_0$, the leading behavior is governed entirely by the linearization of $V(\Delta)$.

Using the auxiliary function $E(\Delta)$ introduced previously, the potential is exactly
\begin{equation}
    V(\Delta) = \Delta^2 \left[-1 + \frac{E(\Delta)}{E(\Delta_0)} \right].
\end{equation}
Since in this regime $\epsilon:=\frac{\Delta-\Delta_0}{\Delta_0}\ll 1$, we write $\Delta=\Delta_0(1+\epsilon)$ and
 Taylor expands $V(\Delta)$ in $\epsilon$. In leading order, we find
\begin{equation}
    V(\Delta) = \Delta_0^2 \frac{(\Delta - \Delta_0)E'(\Delta_0)}{E(\Delta_0)}[1 + o\big(1\big)].
\end{equation}
Substituting this linearized potential into the integral yields
\begin{align}
    t_S &= \int_{\Delta_0}^S \frac{d\Delta}{\sqrt{V(\Delta)}} \nonumber \\
    &\approx \frac{1}{\Delta_0} \sqrt{\frac{E(\Delta_0)}{E'(\Delta_0)}} \int_{\Delta_0}^{S} \frac{d\Delta}{\sqrt{\Delta - \Delta_0}} \nonumber \\
    &= \frac{2}{\Delta_0} \sqrt{\frac{E(\Delta_0)}{E'(\Delta_0)}} \sqrt{S-\Delta_0} + o(1).
\end{align}

To evaluate the prefactor, recall the critical relation $g^{-2} = E(\Delta_0)$. Treating $\Delta_0$ as an implicit function of $g$ and differentiating both sides with respect to $g$ via the chain rule gives
\begin{equation}
    \frac{d}{dg} \big[E(\Delta_0)\big] = \frac{d}{dg} \big( g^{-2} \big) \implies E'(\Delta_0) \frac{d\Delta_0}{dg} = -\frac{2}{g^3}.
\end{equation}
Dividing $E(\Delta_0) = g^{-2}$ by this expression for $E'(\Delta_0)$ yields an exact relationship for the ratio:
\begin{equation}
    \frac{E(\Delta_0)}{E'(\Delta_0)} = \frac{g^{-2}}{-\frac{2}{g^3} \big(\frac{d\Delta_0}{dg}\big)^{-1}} = -\frac{1}{2} g \frac{d\Delta_0}{dg}.
\end{equation}
Given the scaling $\Delta_0 \sim g^2$, the derivative scales as $d\Delta_0/dg \sim g$. Therefore, the ratio scales as $E(\Delta_0)/E'(\Delta_0) \sim g^2$. Since the term $\sqrt{S - \Delta_0}$ depends only on the boundary and is independent of $g$, the overall scaling of $t_S$ to leading order for large $g$ evaluates to
\begin{equation}
    t_S \sim \frac{1}{g^2} \sqrt{g^2} \sim \frac{1}{g}.
\end{equation}

As a concrete example, for a model utilizing an error-function transfer function $\operatorname{Erf}\big(x/(\sqrt{2}\sigma)\big)$, characterized by $S-\Delta_0 = \sigma^2$, direct integration of the standard arcsine covariance in the large-$g$ limit explicitly yields
\begin{equation}
    |\operatorname{Im} t_S| \sim \frac{\sigma\sqrt{2}}{g\sqrt{4/\pi-1}}.
\end{equation}

\subsection{Scaling of the operator growth rate}
\begin{figure}
    \centering
    \includegraphics[width=0.8\linewidth]{\paperroot 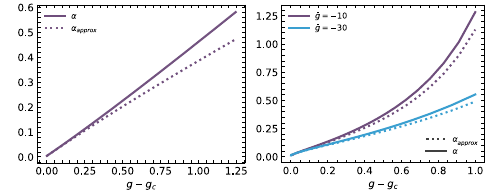}
    \caption{The operator growth rate, $\alpha = \pi/(2t_S)$, vs disorder strength, $g-g_c$. The complex time singularity, $t_S$ is calculated from the exact potential, $V(\Delta)$ (solid line) and approximated from the normal form, $V_p(\Delta)$ for the nonlinearities
    $\operatorname{Erf}(x)$ (\textbf{Left}), and smoothened ReLU (\textbf{Right}). The mean connectivity, $\bar{g}$, is kept constant for the entire range of $g-g_c$ for the smoothened ReLU function.} 
    \label{fig:ts_asymptotics}
\end{figure}

The operator growth rate, $\alpha$, is determined by the imaginary-time singularity via $\alpha = \pi/(2t_s)$. Figure~\ref{fig:ts_asymptotics} illustrates $\alpha$ as a function of the distance to criticality, $g-g_c$. The solid lines represent the exact values of $\alpha$ obtained by numerically integrating the full potential $V(\Delta)$. The dashed lines show the analytical predictions derived from the normal form approximations:
\begin{equation}
    t_s^{(4)} = \tau_{\infty}{\arccos\left(\frac{\Delta_0}{\Delta_0+\sigma^2}\right)} \quad \text{and} \quad t_s^{(3)} =  2\tau_{\infty}\arctan \left(\sqrt{\frac{\sigma^2}{\Delta_0-\Delta_\infty}}\right),
\end{equation}
for the symmetric $\operatorname{Erf}(x/\sqrt{2}\sigma)$ and asymmetric smoothed-ReLU activation functions, respectively. Notably, the departure of the numerical $t_s$ from the leading asymptotic scaling ($\sim \tau_\infty$) occurs more rapidly for the asymmetric case [$\mathcal{O}(\sqrt{\Delta_0})$] compared to the symmetric case [$\mathcal{O}(\Delta_0)$].

We now address the relationship between $\alpha$ and the leading Lyapunov exponent, $\lambda_L$, to verify the bound $\alpha \geq \lambda_L$. While this inequality is visually evident from numerical evaluations deep in the chaotic phase (see Fig.~2 of the main text), the behavior near the critical point warrants explicit analytical verification. Here, we demonstrate that the bound holds strictly for symmetric analytic non-linearities, such as the error function.

For a smooth, odd activation function with a negative third derivative at the origin, $\phi'''(0) < 0$, the chaotic DMFT solution emerges continuously from zero amplitude at the critical point~\cite{Pazo2024}. In this regime, the leading near-critical scaling of the Lyapunov exponent is entirely governed by the local Taylor expansion of $\phi(x)$ near zero~\cite{CrisantiSompolinsky2018,EngelkenWolfAbbott2023}. For $\phi(x) = \operatorname{erf}\big(x/(\sqrt{2}\sigma)\big)$, we have $\phi'''(0) = -\sqrt{2/(\pi\sigma^6)} < 0$, placing this activation function in the same near-critical universality class as $\tanh(x)$. Consequently, as $g \to g_c^+$, the order parameter and the Lyapunov exponent scale as $\Delta_0 \sim (g-g_c)$ and $\lambda_L \sim (g-g_c)^2$, respectively~\cite{CrisantiSompolinsky2018}.

Concurrently, expanding the effective potential near the origin yields the relation $\tau_\infty^{-2}\Delta_0^2 \sim B\Delta_0^4$, where $B = \mathcal{O}(1)$. This dictates the scaling of the prefactor: $\tau_\infty^{-1} \sim \Delta_0 \sim (g-g_c)$. Because the normal form analysis establishes that $\alpha \sim \tau_\infty^{-1}$ near the transition, we find that the operator growth rate scales linearly with the distance to the critical point. 

Comparing the two asymptotic regimes, it follows that sufficiently close to the critical point:
\begin{equation}
    \lambda_L \sim (g-g_c)^2 \ll (g-g_c) \sim \alpha.
\end{equation}
Thus, the quantum bound on chaos is strictly satisfied in the near-critical regime.

\section{WKB analysis to obtain prediction horizon for finite truncation of the Krylov lattice}
The goal of this section is to show that for a finite Krylov lattice, the prediction horizon grows logarithmically with the size of the lattice. We establish this by using the elementary identity that the prediction error is equal to the squared mass of the evolved wave function that has leaked into the unresolved tail. 

Let
\begin{equation}
    |s(0)\rangle = \sum_{k\ge0}s_k(0)|e_k\rangle
\end{equation}
be a white Gaussian initial Krylov state, with $\mathbb{E}[s_k(0)]=0$ and $\mathbb{E}[s_k(0)\overline{s_\ell(0)}]=\delta_{k\ell}$. The observed boundary process is $X(t) := \langle e_0|e^{-iJt}|s(0)\rangle$. We denote the deterministic boundary-to-site propagator by
\begin{equation}
    u_k(t) := \langle e_k|e^{-iJt}|e_0\rangle .
\end{equation}
Since $J$ is a real Jacobi matrix, $e^{-iJt}$ is complex symmetric, and thus $\langle e_0|e^{-iJt}|e_k\rangle = \langle e_k|e^{-iJt}|e_0\rangle = u_k(t)$. Therefore, the boundary process can be written as
\begin{equation}
    X(t) = \sum_{k\ge 0}u_k(t)\,s_k(0).
\end{equation}

Assume that at time $t=0$ we observe only the first $p+1$ Krylov coordinates, $h_p := \bigl(s_0(0),\dots,s_p(0)\bigr)$. The conditional mean of $X(t)$ given the observed state is
\begin{equation}
    \mathbb{E}[X(t)\mid h_p] = \sum_{k\ge0}u_k(t)\,\mathbb{E}[s_k(0)\mid h_p].
\end{equation}
For $k\le p$, we have $\mathbb{E}[s_k(0)\mid h_p]=s_k(0)$, since these coordinates are directly observed. For $k>p$, independence yields $\mathbb{E}[s_k(0)\mid h_p]=0$. Hence,
\begin{equation}
    \mathbb{E}[X(t)\mid h_p] = \sum_{k=0}^p u_k(t)\,s_k(0).
\end{equation}
Let $P_p := \sum_{k=0}^p |e_k\rangle\langle e_k|$ be the orthogonal projection onto the resolved Krylov subspace. The conditional mean can be rewritten compactly as $\mathbb{E}[X(t)\mid h_p] = \langle e_0|e^{-iJt}P_p|s(0)\rangle$. Accordingly, the prediction residual is
\begin{equation}
    X(t)-\mathbb{E}[X(t)\mid h_p] = \langle e_0|e^{-iJt}(I-P_p)|s(0)\rangle = \sum_{k>p}u_k(t)s_k(0).
\end{equation}
Taking the mean-square error and utilizing $\mathbb{E}[s_k(0)\overline{s_\ell(0)}]=\delta_{k\ell}$, we obtain the conditional mean-square prediction error:
\begin{equation}
    \mathcal{E}_p(t) := \mathbb{E}\left[ \left|X(t)-\mathbb{E}[X(t)\mid h_p]\right|^2 \right] = \sum_{k>p}|u_k(t)|^2 .
\end{equation}

Recall from definition~\eqref{eq:sigma_infty} that 
\begin{equation}
    \sigma_{\infty}^{2}(p,\tpred) := \frac{2}{\tpred} \int_{0}^{\infty} \mathcal{E}_p(t) e^{- 2t/\tpred}\,dt.
\end{equation}
Fix a tolerance $\varepsilon\in(0,1)$ and define the prediction horizon $\tpred^*(p,\varepsilon)$ implicitly as the solution to 
\begin{equation}
    \sigma_{\infty}^2(p,\tpred^*(p,\varepsilon)) = \varepsilon.
\end{equation}
In the following, we show that within a semiclassical (WKB) approximation for the amplitudes $u_k(t)$ in the tail ($p\gg 1$), the prediction horizon satisfies $\tpred^*(p,\varepsilon)\sim -(\alpha \log\varepsilon)^{-1}\log p$.

The transition amplitudes satisfy the exact discrete Schrödinger equation
\begin{equation}
    \label{eq:discrete-schrodinger}
    i\,\partial_t u_k(t) = \alpha_{k-1}u_{k-1}(t) + \alpha_k u_{k+1}(t), \qquad u_k(0)=\delta_{k0}.
\end{equation}
For large $k$, the linearly growing hopping coefficients $\alpha_k\sim \alpha k$ vary slowly on the lattice scale. Thus, for $k\gg 1$, the hopping coefficients are locally almost constant in the relative sense:
\begin{equation}
    \frac{|\alpha_{k+1}-\alpha_k|}{\alpha_k} \sim \frac{1}{k} \ll 1.
\end{equation}
The elementary solutions of the corresponding constant-coefficient problem are plane waves. The slow relative variation of $\alpha_k$ suggests patching together such local plane waves with a slowly varying envelope and a local quasimomentum, i.e., using a discrete WKB ansatz:
\begin{equation}
    u_k(t) \approx A(k,t)\exp\bigl(iS(k,t)\bigr),
\end{equation}
where the amplitude envelope $A(k,t)$ varies slowly compared to the phase. At leading order in the gradients of $A$, this yields the eikonal equation
\begin{equation}
    \partial_t S(k,t) + H\bigl(k, \partial_k S(k,t)\bigr) = 0,
\end{equation}
with the effective WKB Hamiltonian $H(k,q) = 2\alpha(k)\cos q$. The associated ray equations (Hamilton–Jacobi characteristics) are
\begin{equation}
    \label{eq:rays}
    \dot{k}(t) = \partial_q H(k,q) = -2\alpha(k)\sin q, \qquad
    \dot{q}(t) = -\partial_k H(k,q) = -2\alpha'(k)\cos q.
\end{equation}
The initial data for the Hamilton--Jacobi problem defines a one-dimensional curve in phase space, which we parametrize as $s\mapsto (k_0(s),q_0(s))$. Therefore, each ray is labeled by $s$.

\medskip\noindent\textit{Position of the front.}---%
The rays with maximal group velocity satisfy $|\sin q|=1$. For propagation towards increasing $k$, we choose the branch $q\approx -\pi/2$, so that $\sin q\approx -1$ and $\dot{k} \approx 2\alpha(k)>0$, while $\cos q\approx 0$ and hence $\dot{q}\approx 0$ along this ray. To describe the leading front propagating towards increasing $k$, we isolate this right-moving branch. Neglecting the residual variation of $q$, we define the approximate WKB front position $k_{\mathrm{front}}(t)$ by
\begin{equation}
    \label{eq:front-ode}
    \frac{d}{dt}k_{\mathrm{front}}(t) \approx 2\alpha\bigl(k_{\mathrm{front}}(t)\bigr) \implies t \approx \int_{k_0}^{k_{\mathrm{front}}(t)} \frac{dk'}{2\alpha(k')}.
\end{equation}
For the linear growth case $\alpha(k) \sim \alpha k$, this yields $k_{\mathrm{front}}(t) \sim e^{2\alpha t}$.

\medskip\noindent\textit{Complex WKB and the dynamic length.}---%
For $k$ in the exponentially small tail beyond this reference scale ($k\gtrsim k_{\mathrm{front}}(t)$), the relevant WKB branch is described by a complex momentum. We write
\begin{equation}
    q(k,t) = \partial_k S(k,t) = q_R(k,t) + i\,q_I(k,t), \qquad q_I(k,t) > 0,
\end{equation}
so that the action acquires a positive imaginary part and the amplitudes decay as
\begin{equation}
    \label{eq:c-tail-WKB}
    |u_k(t)| \sim \exp\left(-\int_{k_{\mathrm{front}}(t)}^{k} q_I\bigl(k',t,s(k',t)\bigr)\,dk'\right),
\end{equation}
where $s(k',t)$ is defined implicitly by $k(t,s(k,t))=k$. The scale $k_{\mathrm{front}}(t)$ serves as the reference front from which we measure the exponential decay into the tail. Writing $k = k_{\mathrm{front}}(t) + \delta k$, we define the local decay length $\ell(t)$ by
\begin{equation}
    \label{eq:dec_len}
    q_I\bigl(k_{\mathrm{front}}(t),t,s(k_{\mathrm{front}},t)\bigr) = \frac{1}{\ell(t)}.
\end{equation}
Expanding $q_I$ locally gives
\begin{equation}
    \int_{k_{\mathrm{front}}(t)}^{k} q_I(k',t)\,dk' \approx \frac{\delta k}{\ell(t)} + \frac{1}{2}\partial_k q_I\bigl(k_{\mathrm{front}}(t),t\bigr)\,\delta k^2 .
\end{equation}
At the WKB level, we assume that $q_I$ varies slowly on the scale of the decay length, such that $|\partial_k q_I|\,\delta k \ll 1/\ell(t)$ in the regime of interest. Thus, the quadratic term is subleading, and the tail assumes the simple exponential form
\begin{equation}
    |u_k(t)| \sim \exp\left(-\frac{k-k_{\mathrm{front}}(t)}{\ell(t)}\right), \qquad
    |u_k(t)|^2 \sim \exp\left(-2\frac{k-k_{\mathrm{front}}(t)}{\ell(t)}\right).
\end{equation}

To relate $\ell(t)$ to $k_{\mathrm{front}}(t)$, we use the conservation of the Hamiltonian. Along a complex ray labeled by its initial data $(k_0,q_0)$, the Hamiltonian is conserved. After analytically continuing the outgoing branch as $q = -\pi/2 + i q_I$, we may write
\begin{equation}
    \alpha\bigl(k(t;s)\bigr)\sinh q_I(t;s) = h(s),
\end{equation}
where $h(s)$ is constant along that ray. Near the front, $q_I\ll 1$, implying $\sinh q_I\simeq q_I$. For asymptotically linear hopping, $\alpha(k)\sim \alpha k$. Hence, along this ray,
\begin{equation}
    q_I(t;s) \simeq \frac{h(s)}{\alpha\bigl(k(t;s)\bigr)} \sim \frac{h(s)}{\alpha\, k(t;s)}.
\end{equation}
Evaluating this on the ray that reaches the front ($s = s_{\mathrm{f}}(t)$), we obtain
\begin{equation}
    \frac{1}{\ell(t)} = q_I\bigl(k_{\mathrm{front}}(t),t\bigr) \sim \frac{h(s_{\mathrm{f}}(t))}{\alpha\,k_{\mathrm{front}}(t)} \implies \ell(t) \sim \frac{\alpha}{h(s_{\mathrm{f}}(t))}\, k_{\mathrm{front}}(t).
\end{equation}
We assume that $h(s_{\mathrm{f}}(t))$ is subleading compared to $k_{\mathrm{front}}(t)$, i.e., $\log h(s_{\mathrm{f}}(t)) = o(\log k_{\mathrm{front}}(t))$. It follows that $\ell(t)$ shares the same leading exponential scaling as $k_{\mathrm{front}}(t)$. This assumption can be verified explicitly in the exactly solvable sech model \cite{Parker}, where:
\begin{equation}
    u_k(t) = (-i)^k \operatorname{sech}(\alpha t) \tanh^k(\alpha t) = \operatorname{sech}(\alpha t) \exp\left[i\left(-\frac{\pi k}{2}-ik\log(\tanh(\alpha t))\right)\right].
\end{equation}
Therefore, $q_I = -\log(\tanh(\alpha t))$, which is independent of $k$. Using the hyperbolic identity, we find
\begin{equation}
    \sinh(q_I(t)) = \sinh\bigl(-\log(\tanh(\alpha t))\bigr) = \frac{1}{2}\bigl(\coth(\alpha t)-\tanh(\alpha t)\bigr) = \frac{1}{\sinh(2\alpha t)}.
\end{equation}
This yields $h(s) = \alpha k(t,s) / \sinh(2\alpha t)$. The ray equation gives $\dot{k} = 2\alpha k \coth(2\alpha t)$, which solves to $k(t,s) = C(s)\sinh(2\alpha t)$. Using the definition of the front, we get
\begin{equation}
    C(s_{\mathrm{front}}) = \frac{k_{\mathrm{front}}(t)}{\sinh(2\alpha t)} = \frac{e^{2\alpha t}}{\sinh(2\alpha t)},
\end{equation}
and thus $h(s_{\mathrm{f}}) \sim \alpha C(s_{\mathrm{front}}) \sim \text{constant}$.

\medskip\noindent\textit{Tail error and prediction horizon.}---%
Using the local form of $|u_k(t)|^2$, we approximate the tail error as
\begin{equation}
    \label{eq:E-tail-WKB}
    \mathcal{E}_p(t) \sim \int_p^\infty \exp\left(-2\frac{k-k_{\mathrm{front}}(t)}{\ell(t)}\right)\,dk = \frac{\ell(t)}{2}\exp\left(-2\frac{p-k_{\mathrm{front}}(t)}{\ell(t)}\right).
\end{equation}
Imposing the instantaneous tolerance $\mathcal{E}_p(t_*(p,\eta)) = \eta$ for $0<\eta<1$, and substituting $\ell(t)\sim C k_{\mathrm{front}}$, we obtain
\begin{equation}
    \frac{2\bigl(p-k_{\mathrm{front}}(t_*)\bigr)}{C k_{\mathrm{front}}(t_*)} \sim \log\left(\frac{C k_{\mathrm{front}}(t_*)}{2\eta}\right) \implies p \sim k_{\mathrm{front}}(t_*) \left[ 1 + \frac{C}{2} \log\left(\frac{C k_{\mathrm{front}}(t_*)}{2\eta}\right) \right].
\end{equation}
Using $k_{\mathrm{front}}(t)\sim e^{2\alpha t}$, we invert this relation to find
\begin{equation}
    t_*(p,\eta) = \frac{1}{2\alpha}\log p + \mathcal{O}(\log\log p),
\end{equation}
where the implicit constant in the $\mathcal{O}(\log\log p)$ term depends on the threshold $\eta$. Thus, the instantaneous threshold $\eta$ affects only subleading terms, not the dominant $\log p$ scaling. 

Finally, approximating the error as a step function $\mathcal{E}_p(t) \approx \Theta\bigl(t-t_*(p,\eta)\bigr)$, the time-averaged integral evaluates to $\sigma_{\infty}^2(p,\tpred) \approx e^{-2t_*/\tpred}$. Imposing $\sigma_{\infty}^2(p,\tpred^*(p,\varepsilon)) = \varepsilon$ gives the prediction horizon:
\begin{equation}
    \tpred^* \sim -\frac{2t_*}{\log \varepsilon} \sim -\frac{\log p}{\alpha\log \varepsilon}.
\end{equation}

\section{Extension to small external white noise}
Consider a network driven by an external white-noise input with variance $\sigma^2$. Authors in~\cite{Helias2018} show that an external white-noise drive appears in the DMFT as a nonzero initial velocity \(\dot\Delta(0^+)\). In the weak-noise regime, we neglect the induced change of the effective potential \(V(\Delta,\Delta_0)\) at leading order and keep only the resulting displacement along the same noiseless orbit.This displacement can be parametrized by the shifted boundary value
\(
\Delta_0 \longmapsto \Delta_0+\delta\Delta_0(\sigma^2,g^2).
\)
 Thus, if \(\Delta_{\mathrm{nl}}(\tau)\) denotes the noiseless solution, then the noisy solution is approximated by
\[
\Delta_\sigma(\tau)
\simeq
\Delta_{\mathrm{nl}}\bigl(|\tau|+\tausigma),
\]
where the shift \(\tausigma \) is fixed by \(\dot\Delta(0^+)\). For illustration, consider the simplified normalized kernel \(\Delta_{\mathrm{nl}}(\tau)=\sech(|\tau|)\).

Treating \(\tau_\sigma\) as a small parameter and using
\(
\tanh|u|=1-e^{-|u|}\sech(|u|),
\)
we find
\begin{equation}
\sech(|u|+\tausigma)
=
(1-\tausigma)\sech(u)
+
\tausigma\,e^{-|u|}\sech^2(u)
+
O(\tausigma^2).
\end{equation}
As \(0<\tausigma\ll 1\), the first-order truncation is a convex combination of the kernels therefore it is a kernel itself. It may therefore be realized as
\begin{equation}
X_t^{(\mathrm{tr})}
=
\sqrt{1-\tausigma}\,Y_t^{(1)}
+
\sqrt{\tausigma}\,Y_t^{(2)},
\end{equation}
Here \(Y_t^{(1)}\) and \(Y_t^{(2)}\) are independent centered Gaussian processes with covariance kernels \(\sech|u|\) and \(e^{-|u|}\sech^2|u|\), respectively. The leading term recovers the noiseless kernel, while the first correction introduces an exponentially damped contribution together with nonlinear terms. Since the undamped kernel \(\sech^2|u|\) is fully predictable and represented by corresponed Jacobi matrix \(J\), the exponential factor can be incorporated by applying the OU damping lift to this state-space model. More generally, let \(C(\tau)=\langle e_0,e^{iJ\tau}e_0\rangle\) be the covariance generated by a self-adjoint Lanczos/Jacobi matrix \(J\). Fix \(\lambda>0\) and define the damped kernel \(\tilde C(\tau)=e^{-\lambda \tau}C(\tau)\) for \(\tau\ge 0\). One can extend the state model from $C(\tau)$ to a state model for \(\tilde C(\tau)\).
Namely consider Ornstein--Uhlenbeck process $X_t$ solving
\begin{equation}
dX_t =-AX_t\,dt+B\,dW_t,
\end{equation}
with $A=(\lambda I - iJ)$ and $B=\sqrt{2\lambda}I$,
where $W_t$ is a vector of independent standard (complex) Wiener processes and $I$ is the identity.
If $X_0$ is chosen from the stationary law (equivalently, $\mathbb{E}[X_0]=0$ and 
$\mathbb{E}[X_0X_0^\dagger]=\Sigma$), then for every $\tau\ge 0$,
\[
\mathbb{E}\left[X_{t+\tau}X_t^\dagger\right]
= e^{-(\lambda I-iJ)\tau}\Sigma,
\qquad\text{and hence}\qquad
\mathbb{E}\left[\overline{(e_0^\dagger X_{t+\tau})}(e_0^\dagger X_t)\right]
= e^{-\lambda\tau} C(\tau).
\]
The stationarity condition is
\begin{equation}
A\Sigma+\Sigma A^\dagger = BB^\dagger.
\end{equation}
Here $J$ is self-adjoint, so $A^\dagger=\lambda I+iJ$ and $BB^\dagger=2\lambda I$. If we set
$\Sigma=I$, then
\begin{equation}
A I + I A^\dagger = (\lambda I-iJ)+(\lambda I+iJ)=2\lambda I=BB^\dagger,
\end{equation}
so $\Sigma=I$ is a stationary covariance.

This gives a perturbative state-space representation of the noisy chaotic kernel. Dynamically, the two independent sources have different roles. The source \(Y^{(1)}\) carries the noiseless chaotic covariance and represents motion along the deterministic attractor. The source \(Y^{(2)}\) carries the exponentially damped correction and represents relaxation back toward the attractor after white noise perturbations.
\section{Prediction for the finite size network}
For a realized trajectory \(X(t)\), we use the Krylov coordinates defined
in the main text,
\begin{equation}
	s_n(t)=\phi_n(i\partial_t)X(t),
	\qquad
	|s(t)\rangle=\sum_{n\geq 0}s_n(t)|e_n\rangle,
\end{equation}
where \(\phi_n(\omega)\) are the orthonormal polynomials associated with
the normalized spectral measure. The trajectory is recovered as
\begin{equation}
	\begin{split}
		X(t)
		=
		\operatorname{Re}\langle e_0|s(t)\rangle 
		=
		\operatorname{Re}\sum_{n=0}^{\infty}
		\langle e_0|e^{-iJt}|e_n\rangle s_n(0) 
		=
		\sum_{n=0}^{\infty}\psi_n(t)z_n(0),
	\end{split}
\end{equation}
where
\begin{equation}
	z_n(t)=i^n s_n(t),
	\qquad
	\psi_n(t)=(-i)^n
	\langle e_0|e^{-iJt}|e_n\rangle.
\end{equation}
For a real process with a symmetric spectral measure, both \(z_n(t)\)
and \(\psi_n(t)\) are real. Prediction therefore requires the basis
functions \(\psi_n(t)\) and an estimate of the real Krylov coordinates
\(z_n(t_0)\) at the forecast origin \(t_0\), obtained from the observed
past.

The functions \(\psi_n(t)\) are calculated from the mean-field
covariance. For the \(\operatorname{erf}\) transfer function, the
normalized covariance
\(
	\rho(t)=\frac{\Delta(t)}{\Delta(0)}
\)
satisfies
\begin{equation}
	\ddot{\rho}(t)
	=
	\rho(t)
	-
	\frac{2g^2}{\pi\Delta(0)}
	\arcsin\!\bigl(\kappa\rho(t)\bigr),
	\qquad
	\kappa
	=
	\frac{2\Delta(0)}{1+2\Delta(0)},
\end{equation}
with stationary initial conditions
\(
	\rho(0)=1\), 
	\(
	\dot{\rho}(0)=0.
\)
We expand the covariance about the origin as
\begin{equation}
	\rho(t)
	=
	\sum_{k=0}^{\infty}
	\frac{\rho^{(2k)}(0)}{(2k)!}t^{2k}.
\end{equation}
Substitution into the mean-field equation and matching equal powers of
\(t\) recursively determine all even derivatives
\(\rho^{(2k)}(0)\). The moments of the normalized spectral measure are
then
\begin{equation}
	\mu_{2k}
	=
	(-1)^k\rho^{(2k)}(0),
	\qquad
	\mu_{2k+1}=0.
\end{equation}

Applying the Stieltjes procedure~\cite{Walter1982}, with polynomial
inner products evaluated algebraically from the moment sequence
$\{\mu_k\}_{k\geq 0}$, yields the Lanczos coefficients
$\{\alpha_n\}_{n\geq 0}$. Equivalently, the procedure orthonormalizes
\(1,\omega,\omega^2,\ldots\) with respect to the normalized spectral
measure. Since the measure is symmetric, the diagonal recurrence
coefficients vanish, and the orthogonal polynomials satisfy
\begin{equation}
	\phi_0(\omega)=1,
	\quad
	\phi_1(\omega)=\frac{\omega}{\alpha_0}, 
    \quad
	\phi_{n+1}(\omega)
	=
	\frac{
		\omega\phi_n(\omega)
		-
		\alpha_{n-1}\phi_{n-1}(\omega)
	}{\alpha_n}.
\end{equation}
The temporal propagation functions are then evaluated as
\begin{equation}
	\psi_n(t)
	=
	(-i)^n\phi_n(i\partial_t)\rho(t).
\end{equation}
The derivatives required to evaluate \(\psi_n(t)\) are generated
recursively by differentiating the mean-field equation.

We next estimate a finite-dimensional Krylov state from a discretely
sampled trajectory obtained by numerically integrating the full
finite-\(N\) RNN dynamics. For a single-neuron trajectory \(X(t)\), we
collect the observations preceding the forecast origin \(t_0\) into
\begin{equation}
	\mathbf{y}
	=
	\left(
	X(t_0),
	X(t_0-\delta t),
	X(t_0-2\delta t),
	\ldots,
	X(t_0-(M-1)\delta t)
	\right)^{\mathsf T},
\end{equation}
where \(\delta t\) is the sampling interval and
\(
	T_{\mathrm{past}}=(M-1)\delta t
\)
is the duration of the observed past window.
After truncating the fitting basis to the first \(K\) Krylov
coordinates, the sampled past satisfies
\begin{equation}
	\mathbf{y}\simeq B\mathbf{z},
\end{equation}
where
\(
	B_{in}
	=
	\psi_n(-\tau_i),
	\)
	\(i=0,\ldots,M-1\)
	\(n=0,\ldots,K-1\)
and
\(
	\mathbf{z}
	=
	\left(
	z_0(t_0),
	z_1(t_0),
	\ldots,
	z_{K-1}(t_0)
	\right)^{\mathsf T}.
\)
We estimate \(\mathbf{z}\) by minimizing the trapezoidally weighted
least-squares error,
\begin{equation}
	\widehat{\mathbf{z}}
	=
	\arg\min_{\mathbf{z}\in\mathbb{R}^K}
	\left\|
	W^{1/2}(B\mathbf{z}-\mathbf{y})
	\right\|_2^2,
\end{equation}
where
\(
	W
	=
	\operatorname{diag}
	\left(
	\frac{1}{2},
	1,
	\ldots,
	1,
	\frac{1}{2}
	\right).
\)
The common quadrature factor \(\delta t\) is omitted because it does not
affect the minimizer. Because \(W^{1/2}B\) is ill-conditioned, we use a
rank-\(r\) truncated singular-value decomposition. 
After estimating \(\widehat{\mathbf{z}}\), we propagate its first \(p\)
Krylov coordinates,
\begin{equation}
	\widehat{X}_p(t_0+\tau)
	=
	\sum_{n=0}^{p-1}
	\psi_n(\tau)\widehat{z}_n(t_0),
	\qquad
	p\leq K.
\end{equation}
The normalized mean-squared prediction error is computed at a fixed
forecast origin \(t_0\) and averaged over the full network,
\begin{equation}
	\mathcal{E}_p(\tau)
	=
	\frac{1}{N\Delta(0)}
	\sum_{a=1}^{N}
	\left[
	X_a(t_0+\tau)
	-
	\widehat{X}_{p,a}(t_0+\tau)
	\right]^2.
\end{equation}

In the numerical experiment, we use
\begin{equation}
	N=10^4,
	\qquad
	M=601,
	\qquad
	\delta t=0.005,
	\qquad
	T_{\mathrm{past}}=3,
\end{equation}
together with
\begin{equation}
	K=100,
	\qquad
	r=30,
	\qquad
	p=20.
\end{equation}
The larger fitting dimension \(K\) allows high-order Krylov content to
be represented by nuisance coordinates rather than being incorrectly
projected onto the first \(p\) coordinates. The rank-\(r\) truncation
removes poorly observable combinations of the \(K\) fitting
coordinates, while only the first \(p\) components of the resulting
estimate are used for future propagation.

The value \(p=20\) was chosen sufficiently large that the prediction is
not limited by the propagation truncation. Empirically, the resulting
finite-network prediction error is comparable to the analytical
prediction error obtained when approximately the first twelve Krylov
coordinates are known exactly. Thus, although \(20\) estimated
coordinates are propagated, the finite-network past provides predictive
information equivalent to approximately 12 exactly known Krylov
coordinates.
\begin{figure}
\includegraphics[width=0.9\linewidth]{\paperroot 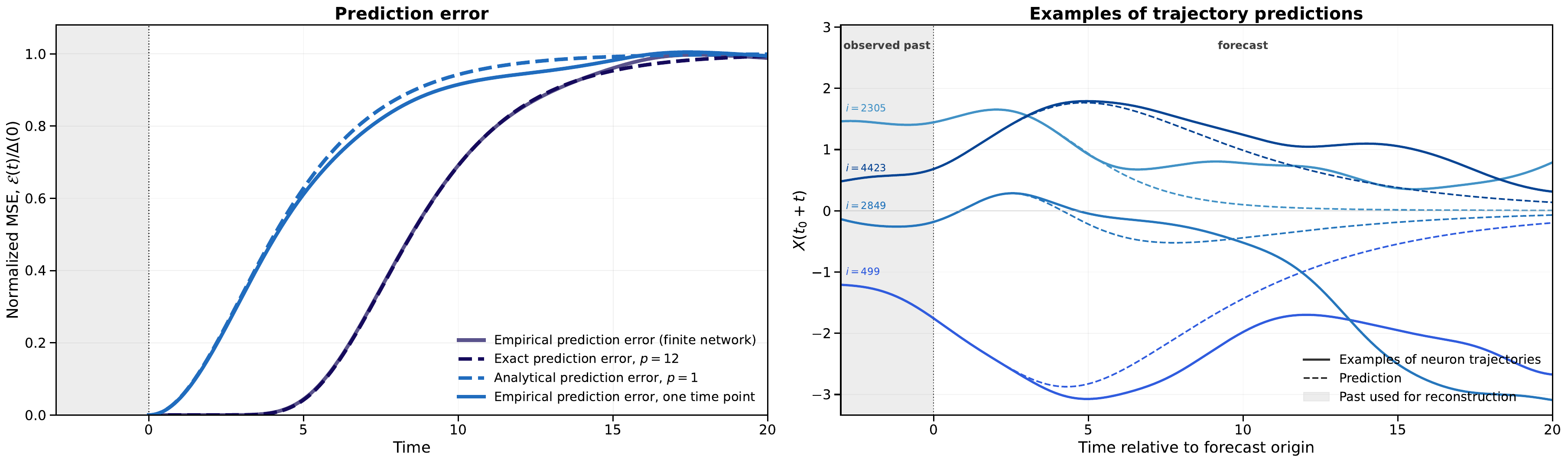}
 \caption{Prediction performance for a finite-size network.
\textbf{Left:} Normalized mean-squared prediction error as a function of time. \textbf{Right:} Representative trajectories of four neurons (indexed by $i$)
and their predicted continuations. Solid lines denote the simulated
trajectories and dashed lines the predictions. The shaded region indicates
the observed past used for reconstruction, and the vertical dotted line at
$t=0$ marks the forecast origin.}
\end{figure}
\clearpage
\bibliography{\paperroot References}

\stopcontents[supp]
\end{document}